\documentclass[apj]{emulateapj}
\usepackage{apjfonts}

\gdef\Ja{$J_1$}
\gdef\Jb{$J_2$}
\gdef\Jc{$J_3$}
\gdef\Ha{$H_1$}
\gdef\Hb{$H_2$}
\gdef\kms{km\,s$^{-1}$}
\gdef\msun{${\rm M}_{\odot}$}
\lefthead{van Dokkum et al.}
\righthead{Growth of Massive Galaxies}
\slugcomment{Accepted for publication in the
Astrophysical Journal}
\begin{document}

\title{The Growth of Massive Galaxies Since $z=2$}

\author{Pieter G.\ van Dokkum\altaffilmark{1,2},
Katherine E.\ Whitaker\altaffilmark{1,2},
Gabriel Brammer\altaffilmark{1,2},
Marijn Franx\altaffilmark{3},
Mariska Kriek\altaffilmark{4,2},
Ivo Labb\'e\altaffilmark{5,2},
Danilo Marchesini\altaffilmark{6,2},
Ryan Quadri\altaffilmark{3,2},
Rachel Bezanson\altaffilmark{1},
Garth D.\ Illingworth\altaffilmark{7},
Adam Muzzin\altaffilmark{1},
Gregory Rudnick\altaffilmark{8,2},
Tomer Tal\altaffilmark{1}, and
David Wake\altaffilmark{9,1}}

\altaffiltext{1}
{Department of Astronomy, Yale University, New Haven, CT 06520-8101}
\altaffiltext{2}{Visiting Astronomer, Kitt Peak National Observatory,
  National O ptical Astronomy Observatory, which is operated by the
  Association of Universiti es for Research in Astronomy (AURA) under
  cooperative agreement with the Nationa l Science Foundation.}
\altaffiltext{3}{Sterrewacht Leiden, Leiden University, NL-2300 RA Leiden,
The Netherlands.}
\altaffiltext{4}{Department of Astrophysical Sciences, Princeton
  University, Princeton, NJ 08544.}
\altaffiltext{5}{Carnegie Observatories, Pasadena, CA 91101.}
\altaffiltext{6}{Department of Physics and Astronomy,
Tufts University, Medford, MA 02155}
\altaffiltext{7}{UCO/Lick Observatory, University of California, Santa
  Cruz, CA 95064.}
\altaffiltext{8}{Department of Physics and Astronomy, University of
  Kansas, Lawrence, KS 66045.}
\altaffiltext{9}{Department of Physics, University of Durham, South Road,
Durham, DH1 3LE, UK}

\begin{abstract}

We study the growth of massive galaxies from $z=2$ to the present
using data from the NEWFIRM Medium Band Survey (NMBS).
The sample is selected at a constant number density of
$n=2\times{}10^{-4}$\,Mpc$^{-3}$,
so that galaxies at different epochs can be compared in a meaningful
way. We show that the stellar mass of galaxies at this number density
has increased by a factor of $\approx{}2$ since $z=2$, following the
relation $\log{}M_n(z)=11.45-0.15z$. In order to determine
at what physical radii this mass growth occurred we construct very
deep stacked rest-frame $R$ band images of galaxies with
masses near $M_n(z)$, at redshifts $\langle{}z{}\rangle{}={}0.6$,
1.1, 1.6, and 2.0. These image stacks of typically 70-80
galaxies enable us to
characterize the stellar distribution to surface brightness limits
of $\sim{}28.5$ mag\,arcsec$^{-2}$. We find
that massive galaxies gradually built up their
outer regions over the past 10 Gyr. The mass within a radius
of $r=5$\,kpc
is nearly constant with redshift whereas the mass at $5\,{\rm kpc}<r<75$\,kpc
has increased by a factor of $\sim{}4$ since $z=2$. Parameterizing
the surface brightness profiles we find that
the effective radius and Sersic $n$ parameter
evolve as $r_e\propto(1+z)^{-1.3}$ and
$n\propto(1+z)^{-1.0}$ respectively.
The data demonstrate
that massive galaxies have grown mostly inside-out,
assembling their extended stellar halos
around compact, dense cores with possibly exponential radial
density distributions. Comparing the observed mass
evolution to the average star formation rates of the galaxies
we find that the growth is likely dominated by
mergers, as in-situ star formation
can only account for $\sim{}20$\,\% of the mass build-up
from $z=2$ to $z=0$.
A direct consequence of these
results is that massive galaxies do not evolve in a self-similar
way: their structural profiles change as a function of redshift,
complicating analyses which (often implicitly) assume self-similarity.
The main uncertainties in this study are possible redshift-dependent
systematic errors in the total stellar masses and the conversion from
light-weighted to mass-weighted radial profiles.

\end{abstract}


\section{Introduction}

Recent studies have found evidence that the structure
of many massive galaxies
has evolved rapidly over the past $\sim 10$\,Gyr.
Galaxies with stellar masses of $\sim 10^{11}$\,\msun\
at $z=1.5-2.5$ are much more compact than galaxies of similar
mass at $z=0$, particularly those
with the lowest star formation rates ({Daddi} {et~al.} 2005; {Trujillo} {et~al.} 2006, 2007; {Toft} {et~al.} 2007; {Zirm} {et~al.} 2007; {van Dokkum} {et~al.} 2008; {Cimatti} {et~al.} 2008; {van der Wel} {et~al.} 2008; {Franx} {et~al.} 2008; {Buitrago} {et~al.} 2008; Stockton et al.\ 2008;
{Damjanov} {et~al.} 2009; {Williams} {et~al.} 2009).
These findings are remarkable as massive galaxies at $z=0$ form
a very homogeneous population, both in terms of their structure and
their (old) stellar populations. As an example, the intrinsic scatter in
the Fundamental Plane relation ({Djorgovski} \& {Davis} 1987) is estimated
to be $\lesssim 0.05$\,dex for the most massive galaxies
(e.g., {Hyde} \& {Bernardi} 2009; {Gargiulo} {et~al.} 2009), which seems difficult to
reconcile with the dramatic changes implied by the measurements
at $z\sim 2$.

Various interpretations of the high redshift data have been offered.
Physical explanations for the
apparent evolution from $z=2$ to $z=0$ include
dramatic mass loss ({Fan} {et~al.} 2008), (minor) mergers
({Naab} {et~al.} 2007; {Naab}, {Johansson}, \& {Ostriker} 2009; {Bezanson} {et~al.} 2009), a fading merger-induced
star burst ({Hopkins} {et~al.} 2009c),
and a combination of
selection effects and mergers ({van der Wel} {et~al.} 2009).
All these models have some observational support, but it is not yet
clear whether any single model is currently capable of simultaneously
explaining the properties of galaxies at $z=2$ and at $z=0$.

The simplest explanation
is that the data are interpreted incorrectly, due to
errors in photometric redshifts,
the conversion from light to stellar mass, the conversion from
light-weighted to mass-weighted radii, or other effects.
It is well known that absolute
mass measurements of distant galaxies are very difficult, even
with excellent data (see, e.g., Muzzin et al.\ 2009a,b for an extended
discussion). Furthermore, sizes are typically determined from
data that do not sample the profiles much beyond the effective
radius $r_e$ (see, e.g., Hopkins et al.\ 2009a, Mancini et al.\ 2009),
even though this is where most of the evolution may have taken place
(e.g., Bezanson et al.\ 2009, Naab et al.\ 2009). Size
measurements also require self-consistent procedures as a function
of redshift, such as analyzing data in the same redshifted
bandpass. It is easier to analyze imaging data in the rest-frame
ultra-violet than in the rest-frame optical at high redshift
(see, e.g., Trujillo et al.\ 2007, Mancini et al.\ 2009), but this
requires large and unknown redshift-dependent corrections for
color gradients.  Despite these uncertainties, it is unlikely
that the small sizes of high redshift galaxies can be entirely
explained by errors, particularly given
the consistency between different studies
(see, e.g., {van der Wel} {et~al.} 2008) and the first measurements
of stellar kinematics ({Cenarro} \& {Trujillo} 2009; {van Dokkum}, {Kriek}, \&  {Franx} 2009a; {Cappellari} {et~al.} 2009).
Nevertheless, subtle redshift-dependent
biases are almost certainly present in the current data.

Ideally, we would measure the mass density profiles of galaxies
well beyond $r_e$ for
large and homogeneously selected samples as a function of redshift.
In this paper, we take some steps in this direction by measuring
the average surface brightness profiles of galaxies at $0<z<2$.
We use new data from the NEWFIRM Medium Band
Survey (NMBS), which provides accurate redshifts and deep
photometry over a relatively wide area.
Galaxies are selected
at a constant number density rather than mass,
which allows a more straightforward
comparison of galaxies as a function of redshift than was possible
in previous studies. The surface brightness profiles are measured
from stacked images, which have a depth equivalent to $\sim
3000$\,hrs of exposure time on a 4\,m class telescope. This depth
allows us to trace the surface brightness profiles to $\sim 28.5$
AB mag\,arcsec$^{-2}$, which is (just)
sufficient to determine whether
the outer envelopes of massive galaxies were already in place at
early times.

As we show in this paper, a self-consistent description
of the structural evolution of massive galaxies can be obtained from
sufficiently deep and wide photometric surveys. Additional data and 
models such as those of {Naab} {et~al.} (2009) and {Hopkins} {et~al.} (2009c)
are needed to better understand the physics driving this evolution.
We assume $\Omega_m=0.3$, $\Omega_{\Lambda}=0.7$, and
$H_0 = 70$\,\kms\,Mpc$^{-1}$. These parameters are slightly different
from the WMAP five-year results (Dunkley et al.\ 2009) but allow
for direct comparisons to most other recent studies of high
redshift galaxies.

\section{Sample Selection}

\subsection{The NEWFIRM Medium Band Survey}
\label{nmbs.sec}

The sample is selected from the NMBS,
a moderately wide, moderately deep near-infrared imaging survey
({van Dokkum} {et~al.} 2009b). The survey
uses the NEWFIRM camera on the Kitt Peak 4m telescope. The camera
images a $28' \times 28'$ field with four arrays. The native
pixel size is $0\farcs 4$; in the reduction the data are resampled
to $0\farcs 3$ pixel$^{-1}$. The gaps between the arrays are relatively
small, making the camera very effective for deep imaging of
$0.25$ deg$^2$ fields. We developed a custom filter system for
NEWFIRM, comprising five medium bandwidth filters in the wavelength
range 1\,$\mu$m -- 1.7\,$\mu$m. As shown in
{van Dokkum} {et~al.} (2009b) these filters pinpoint the Balmer and 4000\,\AA\
breaks of galaxies at $1.5<z<3$, providing accurate photometric
redshifts and improved stellar population parameters.
The survey targeted two $28'\times 28'$
fields: a subsection of
the COSMOS field ({Scoville} {et~al.} 2007), and a field
containing part of the AEGIS strip ({Davis} {et~al.} 2007).
Coordinates and other information are given in
{van Dokkum} {et~al.} (2009b). Both fields have
excellent supporting data, including extremely deep optical $ugriz$
data from the CFHT Legacy
Survey\footnote{http://www.cfht.hawaii.edu/Science/CFHLS/} and deep
Spitzer IRAC and MIPS imaging
({Barmby} {et~al.} 2006; {Sanders} {et~al.} 2007). Reduced CFHT mosaics
were kindly provided to us by the CARS team (Erben et al.\ 2009;
Hildebrandt et al.\ 2009).
The NMBS adds six filters: \Ja, \Jb, \Jc, \Ha, \Hb, and $K$.
Filter characteristics and AB zeropoints of the five medium band filters
are given in {van Dokkum} {et~al.} (2009b).

The NMBS is an NOAO Survey Program, with 45 nights allocated
over three semesters (2008A, 2008B, 2009A). An additional 30 nights
were allocated through a Yale-NOAO time trade.
The data reduction, analysis, and properties of the catalogs are
described in K.\ Whitaker et al., in preparation. In the present study
we use a $K$-selected catalog based on data obtained in semesters 2008A
and 2008B (version 3.1).
The seeing in the combined images is $\approx 1\farcs 1$.
All optical and near-IR images were convolved to the same point-spread
function (PSF) before doing aperture photometry. The analysis in this
paper is based on these PSF-matched images in order
to limit bandpass-dependent
effects.  We note that not much could be gained by using the
original images as the image quality varies
only slightly between the different NEWFIRM bands.
Following previous studies ({Labb{\' e}} {et~al.} 2003; {Quadri} {et~al.} 2007) photometry
was performed in relatively small ``color'' apertures which optimize
the S/N ratio. Total magnitudes in each band
were determined from an aperture correction computed from the $K$
band data. The aperture correction is a combination of the ratio
of the flux in
SExtractor's AUTO aperture (Bertin \& Arnouts 1996)
to the flux in the color aperture and a point-source based correction
for flux outside of the AUTO aperture. We
will return to this in \S\,\ref{select.sec}.

Photometric redshifts were determined with the EAZY code
({Brammer}, {van Dokkum}, \& {Coppi} 2008), using the full $u - 8\mu$m spectral energy distributions
(SEDs) ($u - K$ for objects
in the $\sim 50$\,\% 
of our AEGIS field that does not have Spitzer coverage).
Publicly available redshifts in the COSMOS
and AEGIS fields indicate that the redshift errors are very small
at $\sigma_z/(1+z)<0.02$ (see Brammer et al.\ 2009). Although
there are very few spectroscopic redshifts of
optically-faint $K$-selected galaxies in these fields, we note that
we found a similarly small scatter in a pilot program targeting
galaxies from the {Kriek} {et~al.} (2008) near-IR spectroscopic sample
(see {van Dokkum} {et~al.} 2009b).

Stellar
masses and other stellar population parameters were determined with
FAST ({Kriek} {et~al.} 2009a), using the models of {Maraston} (2005),
the {Calzetti} {et~al.} (2000) reddening law, and exponentially declining
star formation histories.
Masses and star formation rates are based on a
{Kroupa} (2001) initial mass function (IMF);
following {Brammer} {et~al.} (2008)
rest-frame near-IR wavelengths are
downweighted in the fit as their interpretation is
uncertain (see, e.g., {van der Wel} {et~al.} 2006).
Rest-frame $U-V$ colors were measured using the best-fitting EAZY
templates, as described in Brammer et al.\ (2009).
More details are provided in Brammer et al.\
(2009) and, in particular, in K.\ Whitaker et al., in preperation.

\subsection{A Number-Density Selected Sample}
\label{select.sec}

In many studies of galaxy formation and evolution changes in the
galaxy population are traced through the evolution of scaling relations,
such as the Fundamental Plane (see, e.g., {van Dokkum} \& {van der Marel} 2007),
the color-magnitude or color-mass relation (e.g., {Holden} {et~al.} 2004),
and relations between color, size, mass, and surface density
(e.g., {Trujillo} {et~al.} 2007; {Franx} {et~al.} 2008). Other studies focus on evolution
of the luminosity and mass functions, which trace changes in the
number density of galaxies with particular properties
(e.g., {Fontana} {et~al.} 2006; {P{\'e}rez-Gonz{\'a}lez} {et~al.} 2008; {Marchesini} {et~al.} 2009). Finally, some studies
combine information from scaling relations and luminosity functions.
As an example, {Bell} {et~al.} (2004), {Faber} {et~al.} (2007) and others
have inferred significant evolution in the red sequence at $0<z<1$
from the combination of accurate rest-frame colors
and luminosity functions.

\begin{figure}[t]
\epsfxsize=8.5cm
\epsffile{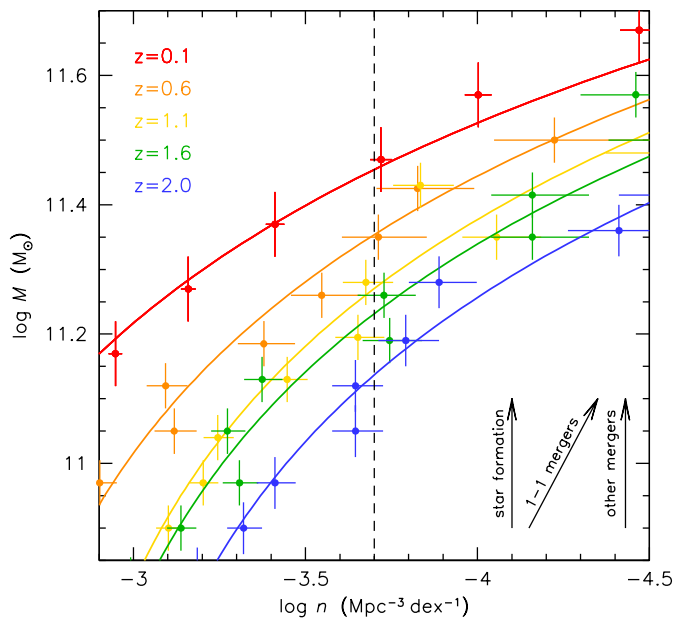}
\caption{\small Evolution of the stellar mass -- number density
relation at $0<z<2$,
derived from Cole et al.\ (2001)
and the NEWFIRM Medium Band Survey data.
Arrows indicate the expected evolution
for star formation, equal mass mergers, and mergers with
mass ratios $<1$. For
most astrophysical processes the most massive
galaxies are expected to evolve along lines of constant
number density, not constant mass. The dashed line shows the
selection applied in this study: a constant number density of
$n = 2 \times 10^{-4}$\,Mpc$^{-3}$.
\label{massfunc.plot}}
\end{figure}
Here we follow a different and complementary approach, selecting
galaxies not by their mass, luminosity, or color but by their number
density. Figure \ref{massfunc.plot} shows stellar mass as a function
of number density (``rotated'' mass functions)
at five different redshifts. The
$z=0.1$ mass function is taken from {Cole} {et~al.} (2001) and converted
to a {Kroupa} (2001) IMF. The points at higher redshift were all
derived from the NMBS data, for $0.2<z<0.8$, $0.8<z<1.4$, $1.4<z<1.8$,
and $1.8<z<2.2$. The datapoints were derived by determining the
number density in bins of stellar mass. No further corrections were
necessary as the completeness
of the NMBS is $\approx 100$\,\% in this mass and
redshift range (see Brammer et al.\ 2009, K.\ Whitaker
et al., in preperation). The data shown in Fig.\
\ref{massfunc.plot} are consistent with those
in {Marchesini} {et~al.} (2009), with smaller (Poisson) errors
due to the much larger area of the NMBS. The lines are simple
exponential fits to the points in the mass range
$10.75<\log M<11.5$;  mass functions from NMBS,
including {Schechter} (1976) fits and a proper error analyis,
will be presented in D.\ Marchesini et al., in preparation.

Arrows indicate schematically how galaxies may be expected to evolve.
Star formation will, to first order,
increase the stellar masses of galaxies and not
change their number density. We note that
this is strictly only true if the
specific star formation rate (sSFR) is independent of mass, which
is in fact not the case (see, e.g., {Zheng} {et~al.} 2007; {Damen} {et~al.} 2009).
Mergers will change both the mass and
the number density. However, because of the steepness of the mass
function in this regime the effect is almost parallel to a line
of constant number density, even for fairly major mergers. This is
demonstrated for mergers with mass ratios $1:10$ -- $1:2$
in Appendix A. We infer that selecting massive galaxies
at a fixed number density enables us to trace the same
population of galaxies through cosmic time, even as they form new stars
and grow through mergers and accretion.
Effectively, we assume that
every massive galaxy today had at least one progenitor at $z=2$ which
was also among the most massive galaxies at that redshift.

\begin{figure}[t]
\epsfxsize=8.5cm
\epsffile{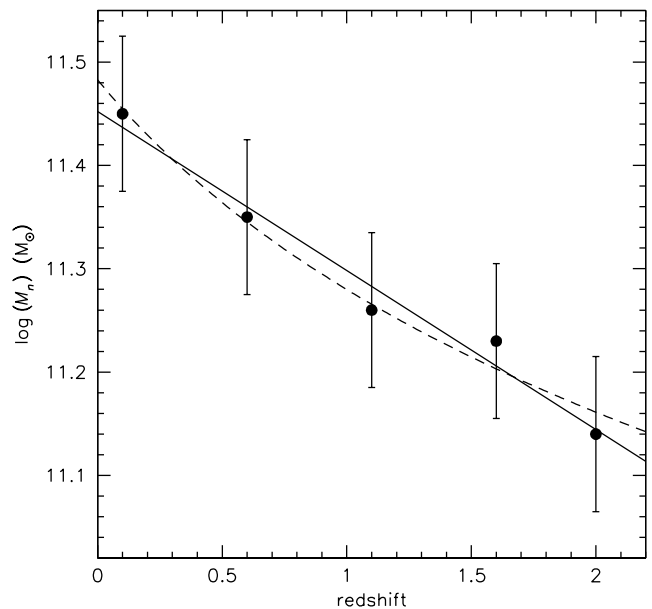}
\caption{\small The stellar mass of galaxies with a number density
of $2 \times 10^{-4}$\,Mpc$^{-3}$, as a function of
redshift. Errorbars are based on estimates of the amount of light
that may be missed in our photometry; random errors are negligible.
The observed
mass evolution is very regular with small scatter.
The solid line is a simple linear fit to the data,
of the form $\log M_{n} = 11.45 - 0.15 z$. The dashed line has
the form $\log M_n = 11.48 - 0.67 \log (1+z)$.
The fits imply that
galaxies with a stellar mass of $3 \times 10^{11}$\,\msun\ today
assembled $\sim 50$\,\% of their mass at $0<z<2$. We note that
unknown systematic uncertainties in the derived stellar masses
have been ignored.
\label{mevo.plot}}
\end{figure}

We choose a number density of $n=2\times 10^{-4}$\,Mpc$^{-3}$
as the selection line in Fig.\ \ref{massfunc.plot}. The choice is
a trade-off between the number of galaxies that enter the analysis
at each redshift, the brightness of these galaxies, and the completeness
of the sample at the highest redshifts.
Figure \ref{mevo.plot} shows the mass evolution of galaxies at this
number density, as given by the intersections of the
exponential fits with the dashed line
in Fig.\ \ref{massfunc.plot}. 
We verified that our results
are not sensitive to the exact number density that is chosen here, by repeating
key parts of
the analysis for a number density of $1 \times 10^{-4}$\,Mpc$^{-3}$.

\begin{figure*}[t]
\epsfxsize=18cm
\epsffile{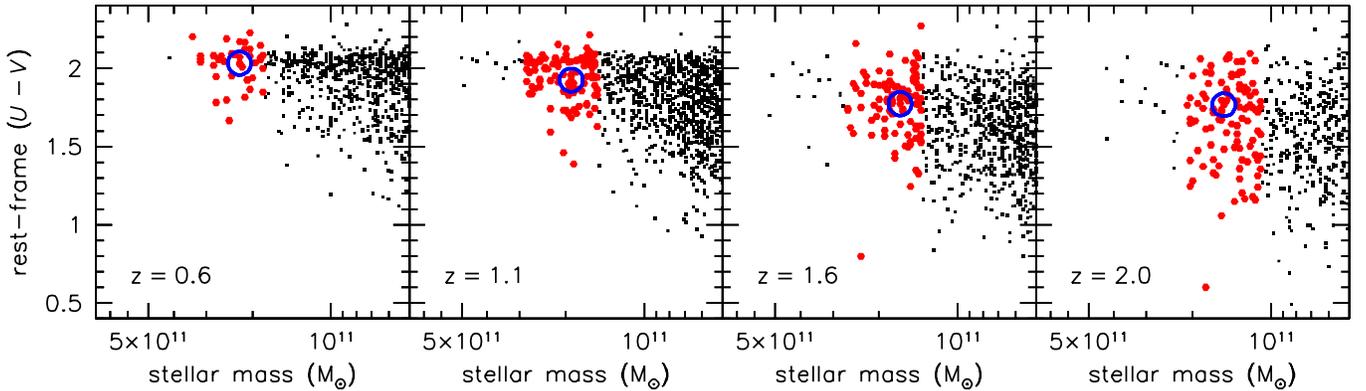}
\caption{\small
Rest-frame $U-V$ color versus mass for galaxies in the
NMBS. 
In each redshift bin galaxies were selected
in a $\pm 0.15$\,dex wide mass bin whose median mass is equal
to $M_n$. Galaxies satisfying this criterion are highlighted
in red. Out to $z\sim 1$ this selection includes mostly
red galaxies. At higher redshifts an increasing fraction of the
sample is blue. This is a real effect, and not due to photometric
errors.
\label{colmass.plot}}
\end{figure*}
The solid line in Fig.\ \ref{mevo.plot} is a simple linear fit to
the data of the form
\begin{equation}
\log M_n = 11.45 - 0.15 z.
\label{mevo.eq}
\end{equation}
The dashed line is an (equally good)
fit of the form $\log M_n = 11.48 -0.67 \log (1+z)$.
Equation \ref{mevo.eq} implies mass growth by a factor of 2 since $z=2$
for galaxies with stellar masses of $3 \times 10^{11}$\,\msun\ today.
The rms scatter in the residuals
is very small at 0.017\,dex, strongly suggesting that
Poisson errors and field-to-field variations are small compared to
other errors.
A potential source of uncertainty is evolution in
the fraction of light that is missed by our photometry. As discussed
by, e.g., {Wake} {et~al.} (2005) and {Brown} {et~al.} (2007), the use of SExtractor's
MAG\_AUTO aperture may lead to biases at faint magnitudes. We
do not use MAG\_AUTO itself
but apply a correction
based on the flux that falls outside the aperture (see {Labb{\' e}} {et~al.} 2003).
This correction is based on pointsources, which means it should be
appropriate in our highest redshift bins where galaxies are small
(see \S\,\ref{stacks.sec}).
The correction may not be appropriate at $z=0.6$ and $z=1.1$, but
at these redshifts the galaxies we select are extremely bright
compared to the limits of our photometry, and the AUTO aperture
is consequently large. From comparing the flux within the AUTO aperture
to the integrated flux of the Sersic fits derived in \S\,\ref{sersic.sec}
we infer that the fraction of flux that is missed ranges from
$\approx 5$\,\% at $z=2$ to $\approx 15$\,\% at $z=0.6$. The
mass evolution from $z=2$ to $z=0.6$
may therefore be slightly underestimated, and
we assign an error of $\pm 0.075$ dex to the mass in
each redshift bin. 

This estimate ignores other
systematic errors in the masses, which
are difficult to assess: uncertainties
in stellar population synthesis codes, the IMF, the treatment of dust,
star formation histories, and metallicities can easily introduce
systematic errors of 0.2 -- 0.3 dex
(see, e.g., {Drory}, {Bender}, \& {Hopp} 2004; {van der Wel} {et~al.} 2006; {Wuyts} {et~al.} 2009; {Muzzin} {et~al.} 2009a; {Marchesini} {et~al.} 2009).
Many of these uncertainties are reduced as we
are only concerned with the relative errors in the masses as
a function of redshift; nevertheless, unknown systematics in the
total masses are probably the largest source
of error in our entire analysis.

\subsection{Properties of the Sample}

In practise, then, we select galaxies with masses near $\log M_n$
in the four
redshift bins that we defined earlier, with $M_n$ given
by Eq.\ \ref{mevo.eq}. The width of each of the mass bins is fixed at
$\pm 0.15$\,dex and the exact bounds are chosen such that the
median mass in the bin is equal to $M_n$. We have
39 galaxies in the $z=0.6$ bin, 108 at $z=1.1$, 96 at $z=1.6$,
and 104 at $z=2.0$. The similarity of the number of
objects in the three highest redshift bins is a reflection of
our selection criterion and the fact that the volumes of these
bins are roughly equal (see Table 1).

Figure \ref{colmass.plot} shows where the selected galaxies
fall in the color-mass plane in each redshift bin. In the lowest
redshift bins galaxies of this number density are nearly always
red, but the range of rest-frame
colors increases as we go to higher redshift.
This increase is real and not due to photometric errors, as
the S/N ratio of the NMBS photometry
is high in this mass and redshift range.
Brammer et al.\ (2009)  use these same data
to demonstrate that the range in colors out to $z=2$
reflects real stellar population differences between the galaxies.
Note that we do not make any cuts on color,
star formation rate, or other properties as we are interested in
the full set of progenitors of today's massive galaxies.

\noindent
\begin{figure}[t]
\epsfxsize=8.5cm
\epsffile{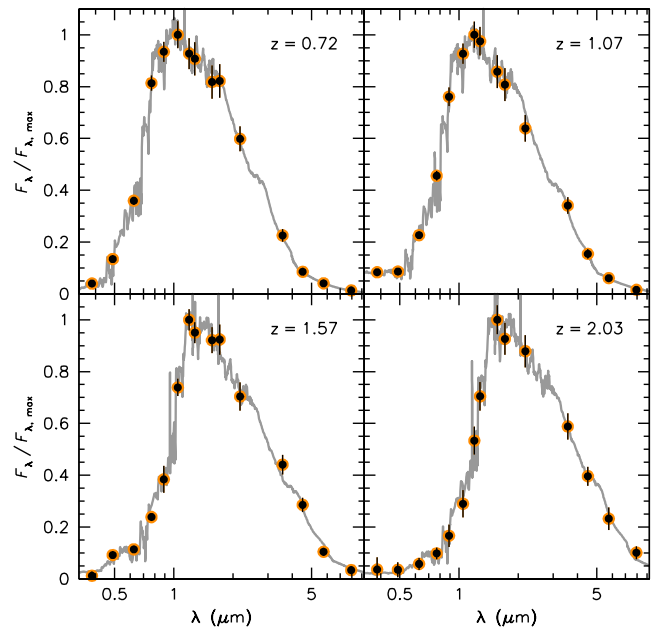}
\caption{\small Spectral energy distributions of typical galaxies
in the four redshift bins, illustrating the high quality of our
photometric data. The locations of these galaxies in the
color-mass plane are indicated by blue circles in Fig.\ \ref{colmass.plot}.
Data points are $u$, $g$, $r$, $i$, $z$
from the Deep CFHT Legacy Survey, \Ja, \Jb, \Jc, \Ha, \Hb, and
$K$ from the NMBS, and IRAC channel 1--4. The grey line shows the
best-fitting EAZY template (Brammer et al.\ 2008). Note that the
medium band filters are critical for determining the redshifts
and SED shapes for galaxies in this mass and redshift range.
\label{seds.plot}}
\end{figure}
The data quality is illustrated in Fig.\ \ref{seds.plot}, which
shows the observed SEDs of four galaxies whose redshifts,
rest-frame
$U-V$ colors and stellar masses are close to the medians in each
redshift bin. The locations of these galaxies in the color -- mass
plane are indicated in Fig.\ \ref{colmass.plot} with blue circles.
The SEDs illustrate the important role of the medium-band near-IR
filters in the analysis; typical massive galaxies at high redshift
are faint in the rest-frame ultra-violet (see also, e.g., {van Dokkum} {et~al.} 2006),
and critical features
for determining redshifts and stellar population parameters are
shifted beyond $\sim 1\,\mu$m. This point was also made
by {Ilbert} {et~al.} (2009), who show that even with 30 photometric bands
(including medium-band optical data from Subaru, but not including
medium near-IR bands) photometric redshifts in the range
$1.5<z<3$ are highly uncertain.

In the present study we are not concerned with (subtle) changes in the
stellar populations of the galaxies as a function of redshift.
Stacked rest-frame SEDs of NMBS galaxies with different
redshifts, masses, and rest-frame colors will be presented in
K.\ Whitaker et al., in preperation. Brammer et al.\ (2009) discuss the
origin of the scatter in the color-magnitude plane,
demonstrating that dusty star-forming galaxies make up most
of the ``green valley'' objects at $0<z<2$. 

\begin{small}
\begin{center}
{{\sc TABLE 1}\\
\sc Properties of Stacked Images}\\
\vspace{0.1cm}
\begin{tabular}{lccccc}
\hline
\hline
 & $z=0$ & $\langle z \rangle=0.6$ & $\langle z \rangle = 1.1$ &$\langle z\rangle=1.6$&$\langle z \rangle = 2.0$ \\
\hline
Source & OBEY & NMBS & NMBS & NMBS & NMBS \\
$z$ range & --- & $0.2-0.8$ & $0.8-1.4$ & $1.4 - 1.8$ & $1.8 -2.2$ \\
$V^{(1)}$ & --- & 0.89 & 2.28 & 1.93 & 2.06 \\
$\log M_n^{(2)}$ & 11.45 & 11.36 & 11.28 & 11.21 & 11.15 \\
$N^{(3)}$ & 14 & 39 & 108 & 96 & 104 \\
$N_{\rm clean}^{(4)}$ & 14 & 32 & 87 & 73 & 79 \\
$r_e^{(5)}$ & $12.4^{1.6}_{-1.3}$&$8.0^{+1.2}_{-0.5}$ & $5.3^{+0.3}_{-0.1}$ & $4.1^{+0.2}_{-0.3}$ & $3.0^{+0.4}_{-0.2}$\\
$n^{(6)}$  & $5.9^{+0.7}_{-0.6}$&$4.0^{+0.4}_{-0.4}$ & $2.9^{+0.2}_{-0.2}$ & $2.5^{+0.2}_{-0.2}$ & $2.1^{+0.5}_{-0.4}$ \\
$\langle {\rm SFR}\rangle^{(7)}$ & --- & $0.8^{+0.3}_{-0.3}$ & $2.5^{+1.1}_{-1.2}$ & $19^{+9}_{-9}$
& $55^{+14}_{-13}$\\
\hline
\end{tabular}
\end{center}
\footnotesize{
(1) Volume in units of $10^6$\,Mpc$^3$.\\
(2) Median of mass bin, in units of \msun. The stacks are normalized such that
$\int_0^{75\,{\rm kpc}} 2 \pi r \Sigma(r) dr = M_n$, with $\Sigma(r)$ the best-fitting Sersic profile.\\
(3) Number of galaxies in mass bins of width $0.3$\,dex. Note that
the densities plotted
in Fig.\ \ref{massfunc.plot} are in units of Mpc$^{-3}$\,dex$^{-1}$.\\
(4) Number of galaxies remaining after visual inspection.\\
arcsec$^{2}$.\\
(5) Best-fitting effective radius in kpc.\\
(6) Best-fitting Sersic (1968) $n$ parameter.\\
(7) Mean star formation rate in units of \msun\,yr$^{-1}$.}
\end{small}

\section{Analysis}
\label{analysis.sec}

\subsection{Creating Stacked Images}

Most studies of the size evolution of distant galaxies
measure effective (i.e., half-light) radii for individual
galaxies and then analyze the evolution of the mean (or median)
size, typically at fixed stellar mass
(e.g., {Trujillo} {et~al.} 2007; {van Dokkum} {et~al.} 2008; {van der Wel} {et~al.} 2008, and many others).
Here we follow a different approach, which emphasizes the strengths of
our dataset: uniform, deep imaging of a large,
objectively defined sample. Instead of measuring sizes and then
taking the average we first create averaged images and then
measure sizes. In Appendix B we show that the average
circularized
effective radius and the Sersic (1968) $n$ parameter can both
be recovered from stacked images of large numbers of galaxies.
The key advantage
of this approach is that it enables the detection of the faint
outer regions of galaxies, which are now thought to evolve much
more strongly than the central regions (e.g., {Naab} {et~al.} 2007, 2009; {Hopkins} {et~al.} 2009c; {Bezanson} {et~al.} 2009).
Rather than parameterizing
structural evolution with changes in $r_e$ only we can characterize
the evolution of the full surface density profiles.
An important practical advantage is that we do not need data of
very high spatial resolution. At $\approx 1\farcs 1$
the resolution of the NEWFIRM data is mediocre even for ground-based
data -- but as we show later this does not prohibit us from tracking
the dramatic changes in galaxy profiles at radii of 5\,kpc -- 50\,kpc.

The stacked images were created by adding normalized, masked
images of the individual galaxies in each redshift bin. Image
``stamps'' of individual objects were cut from the NMBS images.
The stamps are $80 \times 80$ pixels, corresponding to $24\arcsec
\times 24\arcsec$. Images in individual NEWFIRM bands were summed
to increase the S/N ratio. The bands were selected so that the
images are approximately in the same rest-frame band.
Galaxies in the $z=0.6$ redshift bin were taken
from a summed \Ja\,+\,\Jb\ image, galaxies at $z=1.1$ from
\Jc\,+\,\Ha, galaxies at $z=1.6$ from \Ha\,+\,\Hb, and galaxies
at $z=2.0$ from \Hb\,+\,$K$. The corresponding rest-frame
wavelengths are close to the rest-frame $R$ band:
$\lambda_0 = 0.70\,\mu$m, $0.68\,\mu$m,
$0.63\,\mu$m, and $0.65\,\mu$m for $z=0.6$, $z=1.1$, $z=1.6$,
and $z=2.0$ respectively. The galaxies were shifted so that
they are centered as closely as possible
to the center of the central pixel, using
subpixel shifts with a third-order polynomial interpolation.

A mask was created for each object, flagging pixels
that are potentially affected by neighboring galaxies. This mask
image was constructed in the following way. First, SExtractor
was run with a very low detection threshold on a combined
\Jc\,+\,\Ha\,+\,\Hb\,+\,$K$ image. A ``red'' mask
was created by flagging all positive pixels in the segmentation
map except those belonging to the central object. This mask
identifies flux from red objects and bright blue objects but does not
include flux below the detection threshold from the numerous
faint, blue objects that are present in any $24\arcsec \times
24\arcsec$ image of the sky. These objects were identified in
a combined $g + r + i$ image, constructed from the PSF-matched
CFHT Legacy Survey images. These data are extremely deep, reaching
$\approx 29$\,mag (AB) at $5\sigma$ in a $1\farcs 2$ aperture.
With our low detection threshold approximately half of all pixels
are flagged in the blue mask. The final mask is created by
combining the blue and the red mask. The red mask is not redundant,
as a non-negligible number of
objects detected in the NEWFIRM images are absent in the combined
$g+r+i$ image.

The masked images were visually inspected to identify
blended or unmasked objects, star spikes, and other
obvious problems.  This step is necessary as objects that were
flagged as (de-)blended by SExtractor were not removed from the initial
catalogs: given the large size and large apparent brightness
of the galaxies in the lowest redshift bins
a blind rejection
would have introduced redshift-dependent selection effects.
Approximately 25\,\% of objects were removed at this
stage. We verified that the final profiles are not very dependent on
this step; the only individual galaxies which have a significant
impact on the stacks
are the few cases where there are obviously two unmasked objects in
the image.
Next, the images were normalized using the flux in a $10\times 10$ pixel
($3\arcsec \times 3\arcsec$) square aperture. The stacked images
are nearly identical when the catalog
flux is used instead
(in either a fixed aperture or the aperture-corrected flux).
For completeness, the final pre-stack images of all
galaxies are shown in Appendix C.

\noindent
\begin{figure*}[htb]
\epsfxsize=15cm
\epsffile[90 230 446 530]{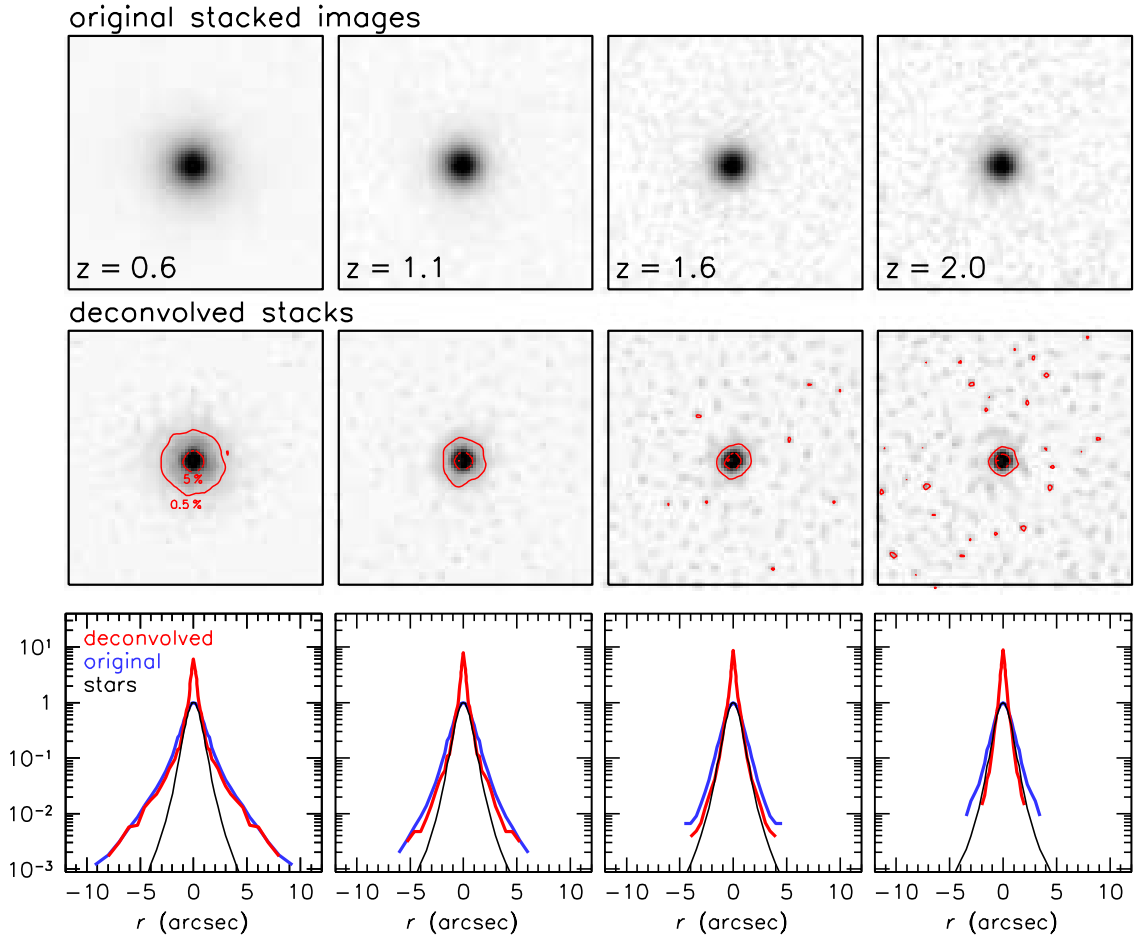}
\caption{\small {\em Top panels:} stacked images of galaxies
with constant number density in four redshift bins. Each image
is $24\arcsec \times 24\arcsec$. The images
reach surface brightness levels of $\sim 28.5$ AB mag arcsec$^{-2}$,
and correspond to $\sim 3000$ hrs of total exposure time on a
4m class telescope. {\em Middle panels:} Deconvolved stacks, highlighting
the fact that the radial extent of the low surface brightness emission
decreases with redshift. Broken (solid)
contours show the radii where the flux is 5\,\% (0.5\,\%) of the
peak flux. The 5\,\% contour is similar at all redshifts,
but the 0.5\,\% contour evolves rapidly. {\em Bottom panels:} Radial
surface brightness profiles, normalized to the peak flux in the
original stacks. Observed profiles are shown in blue, deconvolved
profiles in red. The black curve is for stacked images of stars.
The galaxies are resolved at all redshifts, and are progressively
smaller at higher redshifts.
\label{stacks.plot}}
\end{figure*}
Stacked images were created by summing the individual images.
The masks were also summed, effectively creating a weight map.
Average, exposure-corrected stacked images were created for each
redshift bin by dividing the raw stacks by the weight maps.
The background value at large radii is slightly negative:
the object masks
used in the reduction are not as conservative as the masks used
here, leading to a slight overestimate of the background
in the reduction. Expressed in AB surface brightness
the background error is $\approx 28$ mag\,arcsec$^{-2}$.
We correct for the oversubtracted background in a straightforward
way, by defining
the total flux of a galaxy as the flux within a 75\,kpc radius.
This radius corresponds to $\approx 7 r_e$ for bright
elliptical galaxies at $z=0$, and many tens of effective radii
for high redshift galaxies. In practise, the average value of
pixels with $r>75$\,kpc\ is subtracted from each of the stacks.
This procedure is very robust; bootstrapping the stacks (see
\S\,\ref{densprof.sec}) shows that the uncertainty in the background
correction is only a few percent.
Finally, the images are divided by the total flux in the image.
The final stacks therefore have a total flux of 1 within a 75\,kpc
radius aperture and a mean flux of zero outside of this aperture.

\subsection{Surface Brightness Profiles}
\label{stacks.sec}

The observed stacks are shown in the top panels
of Fig.\ \ref{stacks.plot}. There are no obvious residuals in the
background, thanks to the aggressive masking. The images are very
deep: the surface brightness profiles can be traced to levels of
$\sim 28.5$ AB mag arcsec$^{-2}$ in the observed frame.
For the $z=0.6$ stack these
levels are reached at radii of $\sim 70$\,kpc ($\sim 10\arcsec$);
as we show later this corresponds to $\sim 10$ effective
radii.
The depth
is slightly larger for the $z=0.6$ and $z=1.1$
stacks than for the higher redshift stacks: the $J_x$ band images
are deeper than the $H_x$ and $K$ band data when expressed in AB
magnitudes, and the ellipse fitting routine averages over more
pixels for the low redshift galaxies as they are more extended
(as we show later).

The stellar PSF is fairly broad in this study, with a
full width at half maximum (FWHM) of $\approx 1\farcs 1$, and
we first investigate whether the observed stacks are resolved at
this resolution. Radial surface brightness profiles of the stacked
images are shown in blue in
the bottom panels of Fig.\ \ref{stacks.plot}. Black curves show
the profiles of stacked images of stars, derived from the same
data. The stars were identified based
on their colors (see K.\ Whitaker et al., in preparation) in a
narrow magnitude range similar to the galaxies in the sample.
They were shifted, masked, visually inspected, averaged, and
normalized
in the same way as the galaxy images. The galaxy profiles
and the stellar profiles were normalized to a peak flux of 1.
The blue curves are broader than the black curves at all
redshifts, demonstrating that the galaxies are resolved.

To investigate the behavior of the galaxy profiles with
redshift the stacks were deconvolved using carefully constructed
PSFs. The PSFs were created by averaging images of bright
unsaturated stars, masking companion objects. The COSMOS
and AEGIS fields have slightly different PSFs; for each stack
a separate PSF was constructed using the appropriate filters and
appropriately weighting the PSFs of the two fields. As a test,
we repeated the analysis using the stacked stellar images described
above. Differences were small and not systematic; the differences
in the measured effective radii were $<10$\,\% at all redshifts.
The
deconvolution was done with a combination
of the Lucy-Richardson algorithm ({Lucy} 1974) and
$\sigma$-CLEAN ({H{\"o}gbom} 1974; {Keel} 1991), ensuring
flux conservation.  Lucy works well for extended
low surface brightness emission but does not optimally
recover the flux in the central pixels (see, e.g., {Griffiths} {et~al.} 1994),
whereas CLEAN quickly converges in the central regions but leads to
strong amplification of noise in areas of low surface brightness.
In practice, we applied a smoothly varying weight function to combine
the CLEAN and Lucy reconstructions, giving a weight of 1 to CLEAN in
the central pixels and a weight of 1 to Lucy at radii $>3$ pixels.
In the transition region the
form of the weight
function was determined by the requirement to conserve
total flux. We note that we use the
deconvolved images for illustrative purposes only, as we
later quantify the
evolution by fitting Sersic (1968) profiles to the
original, PSF-convolved images.
The deconvolved images are shown below the original stacks in
Fig.\ \ref{stacks.plot}. Profiles derived from these images
are shown in red in the bottom panels of Fig.\ \ref{stacks.plot}.

It is immediately obvious from the deconvolved images and the
radial profiles that the galaxies are smaller
at higher redshift.\footnote{Note that this trend is somewhat
exaggerated going from $z=0.6$ to $z=1.1$,
as the flux is shown as a function of radius in arcseconds rather
than kpc in Fig.\ \ref{stacks.plot}.}
Furthermore, the central parts of the galaxies
are fairly similar: at all redshifts there is a bright core
but only at lower redshifts this core is surrounded by extended
emission. This is a key result of the paper and it is
quantified in the sections below. Here it is
illustrated by the red contours in
Fig.\ \ref{stacks.plot}. The inner (dotted) contour shows the
radius at which the surface brightness is 5\,\% of the peak
value. This radius is very similar at all redshifts. The outer
(solid) contour shows the radius where the surface brightness
if $0.5$\,\% of the peak. This radius is much larger at low
redshift than at high redshift. Together, the two
contours demonstrate that the
{\em shape} of the profile changes with redshift, with the
core of present-day massive galaxies mostly in place at $z=2$
but the outer parts building up gradually over time.

\subsection{Surface Density Profiles}
\label{densprof.sec}

\begin{figure*}[htbp]
\epsfxsize=14cm
\epsffile[-37 270 470 661]{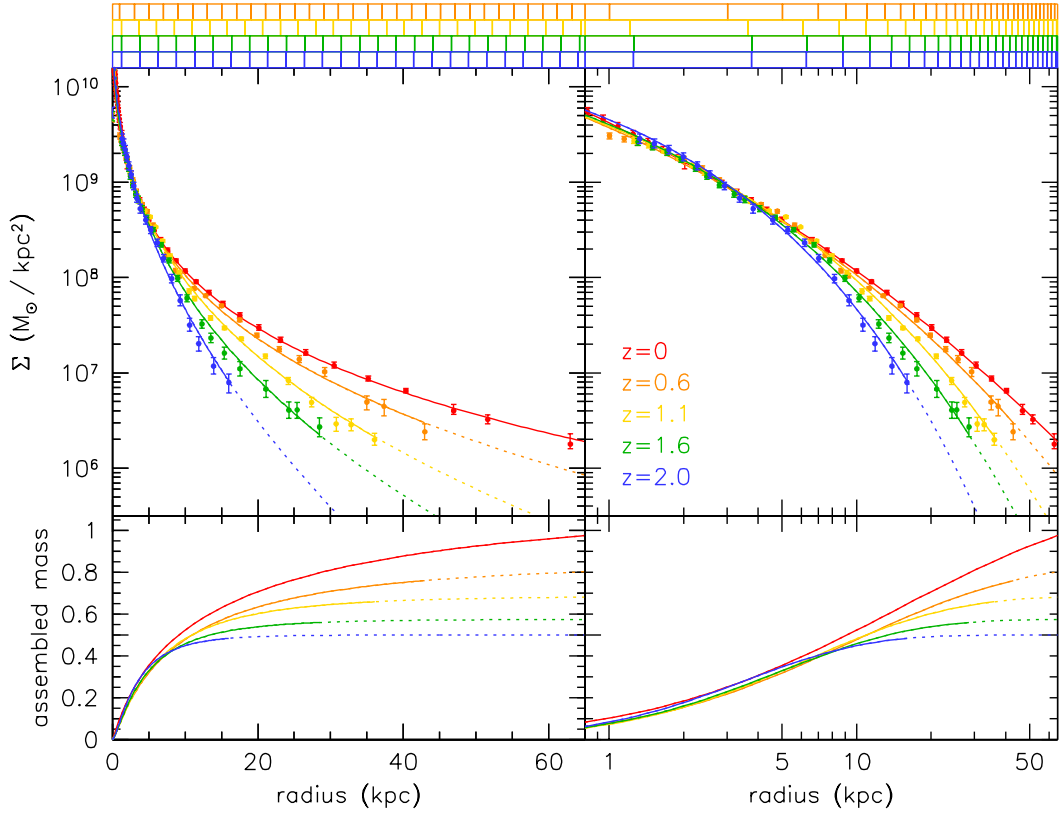}
\caption{\small {\em Top panels:} Average radial surface density profiles
of galaxies with a number density of
$2\times 10^{-4}\,$Mpc$^{-3}$ as a function of redshift.
The data points were measured from the deconvolved stacked images.
Errorbars are 68\,\% confidence
limits derived from bootstrapping the stacks.
The same data are shown versus radius (left
panel) and log radius (right panel). Small boxes above the panels
indicate the pixel size of $0\farcs 3$. 
There is a clear trend
with redshift: at small radii the profiles overlap, but at large radii
the profiles get progressively steeper with redshift.
Lines show the best-fitting Sersic profiles, determined from fitting
PSF-convolved models
to the original (not deconvolved) stacked images.
{\em Bottom panels:} Cumulative mass as a function of radius, as implied
by the best-fitting Sersic profiles. The vertical axis is in units
of the total mass at $z=0$ within a 150\,kpc diameter aperture.
Note that the normalization of the profiles is not a free
parameter but follows from the requirement
that the total mass within this aperture is equal to $M_n(z)$ (Eq.\ 1).
The mass growth of galaxies of this number density is
dominated by the build-up of the outer envelope, at radii
$\gtrsim 5$\,kpc.
\label{growth.plot}}
\end{figure*}

When color gradients are ignored, the deconvolved radial profiles can be
interpreted as stellar mass surface density profiles. The
median mass of the galaxies in each of the stacks is determined
by our constant number density selection, and the calibration of the
profiles follows from the requirement that
\begin{equation}
\int_{0}^{75} 2 \pi r \Sigma(r) dr = M_n,
\label{prof.eq}
\end{equation}
\vspace{0.0cm}\\
with $r$ in kpc,
 $\Sigma(r)$ the radial surface density profile in units
of \msun\,kpc$^{-2}$,
and $M_n$ given by Eq.\,\ref{mevo.eq}.
It is implicitly
assumed that the total stellar mass in our catalog equals the
mass within a 150\,kpc diameter aperture (see \S\,\ref{select.sec}).
Figure \ref{growth.plot} shows the
radial surface density profiles as a function of redshift.
Errorbars are 68\,\% confidence intervals
determined from bootstrapping: 500
realizations were created of each of the stacks, and we followed
the same analysis steps on these as for the actual stacks.
This method is more robust than a formal analysis of the noise,
as it includes errors due to improper masking of particular objects,
uncertainties in the background subtraction, and uncertainties due to
real variation in the properties of galaxies that enter the stack.
We note here that color gradients are almost certainly important (see
\S\,\ref{colgrad.sec}), but that it is at present
difficult to correct for them.

\noindent
\noindent
\begin{figure*}[htbp]
\epsfxsize=15cm
\epsffile[-14 441 515 692]{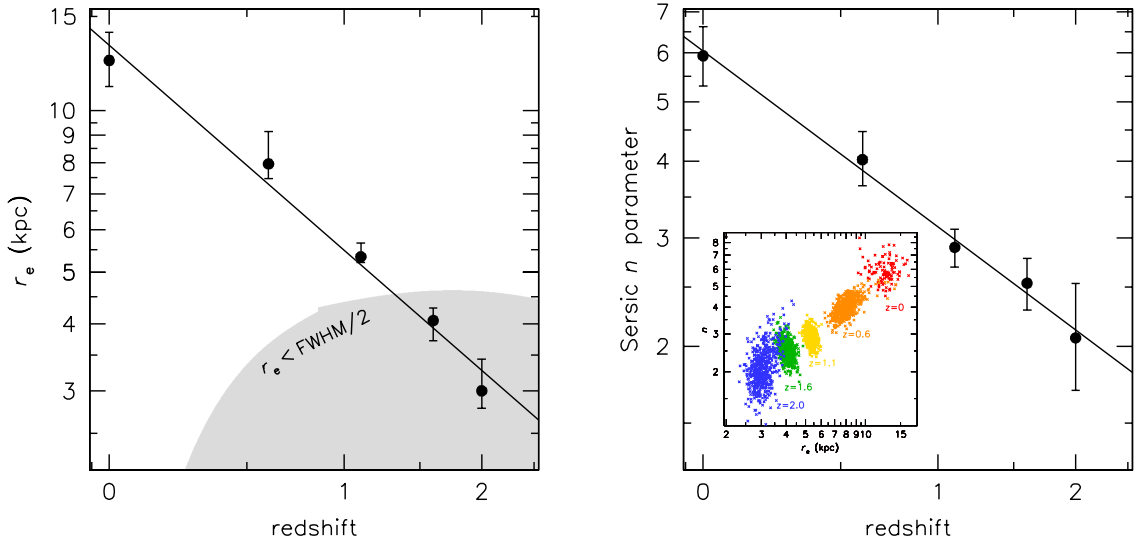}
\caption{\small Evolution of the effective radius $r_e$ {\em (left panel)}
and the Sersic parameter $n$ {\em (right panel)} for galaxies with
a number density of $2 \times 10^{-4}$\,Mpc$^{-3}$.
Errors are 68\,\% confidence intervals determined from repeating
the analysis on bootstrapped realizations of the stacked images.
Individual measurements from these realizations are shown in the
inset.
The grey area indicates where the effective diameter
is smaller than the FWHM of the PSF.
Galaxies have smaller effective radii at higher redshift and
profiles that are closer to exponential. 
\label{sersic.plot}}
\end{figure*}
The profile for $z=0$ was determined from the Observations of Bright
Ellipticals at Yale (OBEY) survey (Tal et al.\ 2009). This survey
obtained surface photometry  out to very large radii
for a volume-limited sample of luminous elliptical galaxies.
A stacked image was created and analyzed in the same way
as was done for the NMBS galaxies; details are given in
Appendix D. As discussed in the Appendix, the OBEY $z=0$ stacked
image should be directly comparable to the NMBS stacks at higher
redshift. Also, its surface density profile was normalized
using Eq.\ \ref{prof.eq} and is therefore on the
exact same system as the NMBS galaxies.

The surface density profiles display a striking evolution with redshift.
At $z=0$, the profile shows the dense center and extended
outer envelope familiar from numerous studies of elliptical galaxies.
At higher redshift, the profiles in the central regions remain
virtually unchanged
but they become progressively steeper at large
radii. The extended outer envelope of elliptical galaxies appears
to have been built up gradually since $z=2$ around a compact core
that was formed at higher redshift. Our data obviously lack the
resolution to properly determine the shape of the
profiles in the central 5\,kpc; nevertheless, flux conservation
implies that they cannot be significantly steeper or flatter
than what is shown in Fig.\ \ref{growth.plot}. More to the
point, the data do have sufficient depth and resolution to track
the emergence of the outer envelope at radii $>5$\,kpc,
although even deeper data would be valuable at $z=2$.
A possible concern is that subtle redshift-dependent effects drive
(part of) the evolution at large radii. We tested this explicitly in
Appendix B, where we redshift the
$z=0$ and $z=0.6$ data to
$z=2$ and show that the derived evolution is robust.

\subsection{Sersic Fits}
\label{sersic.sec}

The profiles are parameterized with standard {Sersic} (1968) fits,
of the form
\begin{equation}
\Sigma_b(r) = \Sigma_e 10^{-b_n[(r/r_e)^{1/n}-1]},
\end{equation}
where $\Sigma(r)$ is the surface brightness at radius $r$,
$b_n$ is a constant that depends on $n$, $n$ is the ``Sersic
index'', and $r_e$ is the radius containing 50\,\% of the
light. These fits are performed on the original stacked images, by fitting
models convolved with the PSF. This approach has the advantage that
it uses a convolution rather than a deconvolution. The fits were
done with GALFIT ({Peng} {et~al.} 2002). They converged quickly, and
the parameters do not depend on the choice of fitting
region, initial guesses for the parameters, and whether the sky
is left as a free parameter. The fits were normalized using
Eq.\ \ref{prof.eq} and therefore give the correct masses within
a 150\,kpc diameter aperture.

The Sersic fits are shown by the lines in the top panels of
Fig.\ \ref{growth.plot}. The lines follow the datapoints quite
well, indicating that the deconvolutions did not produce
large systematic errors in the profiles. The bottom panels
of Fig.\ \ref{growth.plot} show the cumulative radial mass
profiles as implied by the  Sersic fits. The vertical axis
is in units of the total mass within a 150\,kpc diameter
aperture at $z=0$, i.e., $2.8 \times 10^{11}$\,\msun.
The mass contained within $\sim 5$\,kpc is remarkably similar
at all redshifts, and essentially all the mass growth is at
large radii.

The evolution in the shape of the radial surface density
profiles is parameterized by evolution in the effective radius
and in the Sersic parameter $n$. The profiles are
both more concentrated and closer to exponential at redshifts
$z>1.5$. This is demonstrated in Fig.\ \ref{sersic.plot},
which shows the evolution in $r_e$ and $n$. Errorbars are
68\,\% confidence limits determined from bootstrapping the
stacks.
We note that our fitting procedure, and particularly the definition
of total mass (Eq.\ 2), leads to subtle and redshift-dependent
correlations of the errors. The inset in Fig.\ 7 shows individual
measurements of $r_e$ and $n$ from the bootstrapped stacks. Correlations
exist but they are not sufficiently large to influence our results.
The lines are fits to the data of the form
\begin{equation}
\label{re.eq}
r_e = 13.2 \times (1+z)^{-1.27}
\end{equation}
and
\begin{equation}
\label{n.eq}
n = 6.0 \times (1+z)^{-0.95}.
\end{equation}
The formal errors in these relations are small and the scatter
in the residuals is small: 0.029 in $\log r_e$, and $0.015$ in $\log n$.
Together with Eqs.\ \ref{mevo.eq} and Eq.\ \ref{prof.eq} these
expressions provide a complete description of the evolution of
the stellar mass in galaxies with a number density of
$2\times 10^{-4}$\,Mpc$^{-3}$, as a
function of redshift and radius.

\begin{figure}[htb]
\epsfxsize=8.5cm
\epsffile{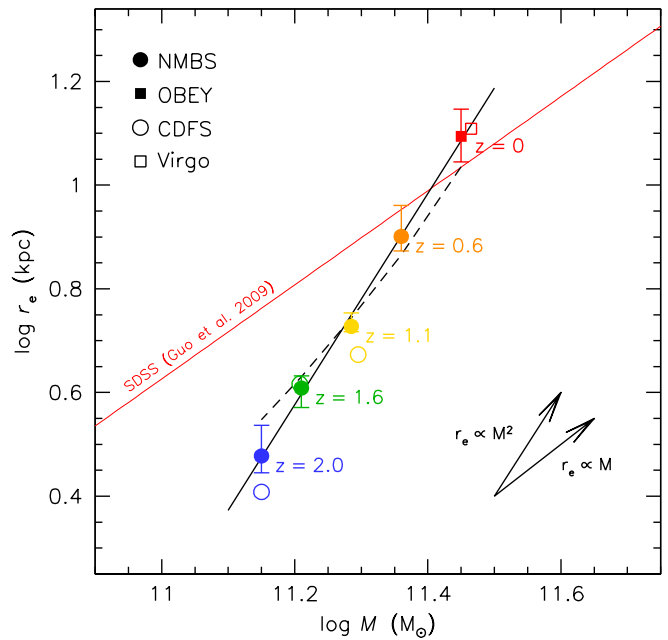}
\caption{\small Evolution in the radius-mass plane.
Our data are consistent with measurements
for individual galaxies of the same masses and redshifts in the
FIREWORKS
CDF-South survey of Wuyts et al.\ (2008) and Franx et al.\ (2008)
(open circles). Our $z=0$ point from the OBEY survey (Tal et al.\ 2009)
is consistent with data from Virgo ellipticals by Kormendy et al.\ (2009)
and a recent determination of the mass-size relation in the SDSS
(Guo et al.\ 2009).  The evolution
in effective radius is stronger than in mass: the solid line is a fit
of the form $r_e \propto M^{2.04}$.
The dashed line is the expected evolution
of the effective radius for inside-out growth, calculated using
Eq.\ \ref{drdmfit.eq} and the measured value of the Sersic
index $n$ at each redshift.
\label{massradius.plot}}
\end{figure}

The evolution in the effective radius is a factor of $\sim 4$,
whereas the mass evolves by a factor of $\sim 2$. 
The evolution
in the familiar radius-mass diagram (see, e.g., {Trujillo} {et~al.} 2007)
is shown in Fig.\ \ref{massradius.plot}. The solid line is a fit
to the OBEY and NMBS data; the slope implies that $r_e \propto M^{2.04}$.
In addition to the OBEY data we show the mass-size
relation for massive early-type galaxies
from {Guo} {et~al.} (2009) (SDSS) and the average of four
Virgo ellipticals from {Kormendy} {et~al.} (2009)
(see Appendix D). The $z=0$
data are in good agreement with each other and also with an extrapolation
of the NMBS data to lower redshift. Open circles  show
the median sizes of galaxies in the GOODS CDF-South field, as determined
by the FIREWORKS survey ({Wuyts} {et~al.} 2008; {Franx} {et~al.} 2008). The CDF-South is
a much smaller field (by a factor of $>10$), but the imaging data
is of very high quality (see {Franx} {et~al.} 2008).
The CDF-South data are in excellent agreement with our results,
although we note that the uncertainties
are large as there are only 10--15 galaxies in each of
the bins. Finally, we note that the sizes of the $z=2$ galaxies
are a factor of $\sim 3$ larger than the median of
nine quiescent galaxies at $z=2.3$ ({van Dokkum} {et~al.} 2008). The
reason is that we include all galaxies in the analysis, not just
quiescent ones, and as is well known
star forming galaxies are significantly larger than quiescent galaxies
(e.g., {Toft} {et~al.} 2007; {Zirm} {et~al.} 2007; {Franx} {et~al.} 2008,
Kriek et al.\ 2009b).

\noindent

\section{Discussion}

\subsection{Inside-Out Growth}
\label{inout.sec}

As demonstrated in \S\,\ref{select.sec} and
\S\,\ref{analysis.sec} galaxies with a space density
of $2 \times 10^{-4}$\,\msun\,Mpc$^{-3}$
increased their mass by
a factor of $\approx 2$ since $z=2$, apparently mostly by adding stars at
large radii. The radial dependence of the evolution
can be assessed by integrating the deprojected density profiles of
the galaxies. Following {Ciotti} (1991) the surface density
profiles were converted to mass density profiles using an Abel
transformation. The mass in the central regions can then be
determined by
integrating these mass density profiles from zero to a fixed
physical radius (see {Bezanson} {et~al.} 2009). {Bezanson} {et~al.} (2009)
used a radius of 1\,kpc, which corresponds to the typical
effective radii of quiescent galaxies at $z\sim 2.3$.
In our data
1\,kpc corresponds to a small fraction of a single pixel, and
we use a fixed radius of 5\,kpc instead.

\begin{figure}[tb]
\epsfxsize=8.5cm
\epsffile{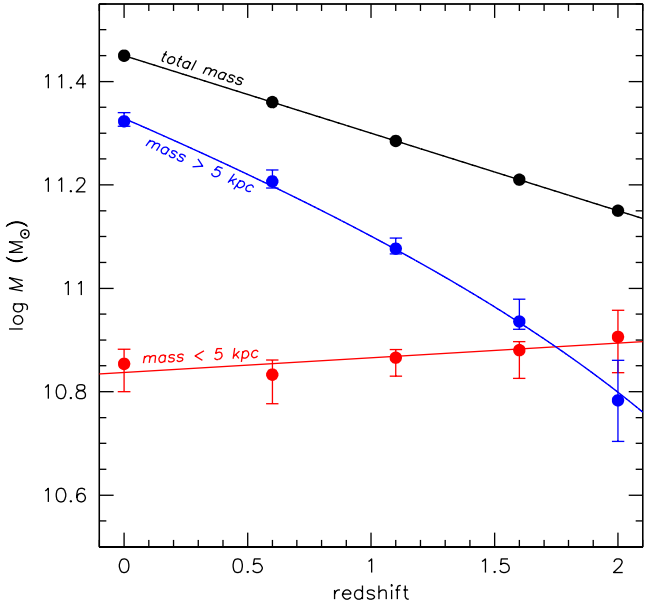}
\caption{\small Comparison of the mass contained within a fixed
radius of 5\,kpc (red curve) to the mass at larger radii (blue curve),
as a function of redshift. Errorbars are 95\,\% confidence limits
derived from bootstrapping. The total mass is shown in black. Galaxies
with number density $n=2 \times 10^{-4}$\,Mpc$^{-3}$ have
a nearly constant mass in the central regions. The factor of $\approx 2$
increase in total mass since $z=2$ is driven by the addition
of stars at radii $>5$\,kpc.
\label{inout.plot}}
\end{figure}

The evolution of the mass within 5\,kpc is shown in
Fig.\ \ref{inout.plot} by the red datapoints. Errors
were determined from 500 bootstrapped realizations of
the stacks. Also shown are the evolution of the total
mass and the evolution of the mass outside a fixed radius
of 5\,kpc.
Note that each of the stacks is normalized to give
exactly the total mass of Eq.\ \ref{mevo.eq}; the total mass
has therefore no errorbar in Fig.\ \ref{inout.plot} and the
errorbars on the red and blue data points are directly coupled.
The mass within a fixed aperture of 5\,kpc is approximately constant with
redshift at $\approx 10^{10.9}$\,\msun, whereas the
mass at $r>5$\,kpc has increased by a factor of $\sim 4$ since $z=2$.

\noindent
It is interesting to consider the expected evolution 
of galaxies in the
radius-mass diagram (Fig.\ \ref{massradius.plot}) in this context.
As discussed in, e.g., {Bezanson} {et~al.} (2009) and {Naab} {et~al.} (2009),
the change in radius for a given change in mass provides important
information on the physical mechanism for growth. Major mergers
are expected to result in a roughly linear relation,
$d\log(r_e)/d\log(M) \sim 1$, whereas minor
mergers could give values closer to 2.
There is, however, also a simple geometrical effect resulting from the
shape of the Sersic profile and the definition of the effective
radius. If mass is added to a galaxy the effective
radius has to change so that it still encompasses 50\,\% of
the total mass.
If the added mass is small and at $r\gg r_e$ the form of the density
profile at $r\approx r_e$ will not change appreciably, even in
projection. The change in effective radius for a given change in
mass is then simply the inverse of the
derivative of the enclosed mass profile,
\begin{equation}
\frac{d\log(r)}{d\log(M)} = \left\{ \frac{d \log \left[ \int_0^{r} 2 \pi r
\Sigma(r) dr \right]}{d\log r}  \right\}^{-1},
\label{drdm.eq}
\end{equation}
evaluated at $r=r_e$. Numerically solving Eq.\ \ref{drdm.eq} gives a simple
relation between the Sersic index $n$ and the change in effective radius
for a given change in mass:
\begin{equation}
\frac{d\log(r_e)}{d\log (M)} \approx 3.56 \log(n+3.09) -1.22.
\label{drdmfit.eq}
\end{equation}
This relation is accurate to $0.01$ dex for $1\leq n \leq 6$.

Equation
\ref{drdmfit.eq} implies that the effective radius increases approximately
linearly with mass if the projected density follows an
exponential profile, but as $r_e \propto M^{1.8}$ for a de Veaucouleurs
profile with $n=4$. This in turn implies that strong evolution in
the measured projected effective radius can be expected in all inside-out
growth scenarios irrespective of the physical mechanism that is
responsible for that
growth, unless the projected density profiles are close to exponential.
The predicted change in $r_e$ as a function of
mass based on Eq.\ \ref{drdmfit.eq}
is indicated with a dashed line in Fig.\
\ref{massradius.plot}, calculated
using the measured values of $n$ at each redshift.
As might have been
expected, the line closely follows the observed data points.

\subsection{Star Formation versus Mergers}

Several mechanisms have been proposed to explain the growth of massive
galaxies. The simplest is star formation, which can be expected
to play an important role at higher redshifts as a large fraction
of massive galaxies at $z\sim 2$ have high star formation rates
(e.g., {van Dokkum} {et~al.} 2004; {Papovich} {et~al.} 2006). {Franx} {et~al.} (2008) expressed
the evolution in terms of surface density, and found that
many galaxies with the (high) surface densities of
$z=0$ early-type galaxies were forming stars at $z=1-2$.
However, the old stellar ages of the most massive early-type galaxies
(e.g., {Thomas} {et~al.} 2005; {van Dokkum} \& {van der Marel} 2007) and the existence of
apparently ``red and dead'' galaxies with small sizes at $z=1.5-2.5$
(e.g., {Cimatti} {et~al.} 2008; {van Dokkum} {et~al.} 2008) suggest that at least some
of the growth is due to (``dry'') mergers. Growth by mergers is
expected in $\Lambda$CDM galaxy formation models
(e.g., {De Lucia} {et~al.} 2006), and could be effective in growing
the outer envelope of elliptical galaxies
({Naab} {et~al.} 2007, 2009; {Bezanson} {et~al.} 2009).

\begin{figure}[tb]
\epsfxsize=8.5cm
\epsffile{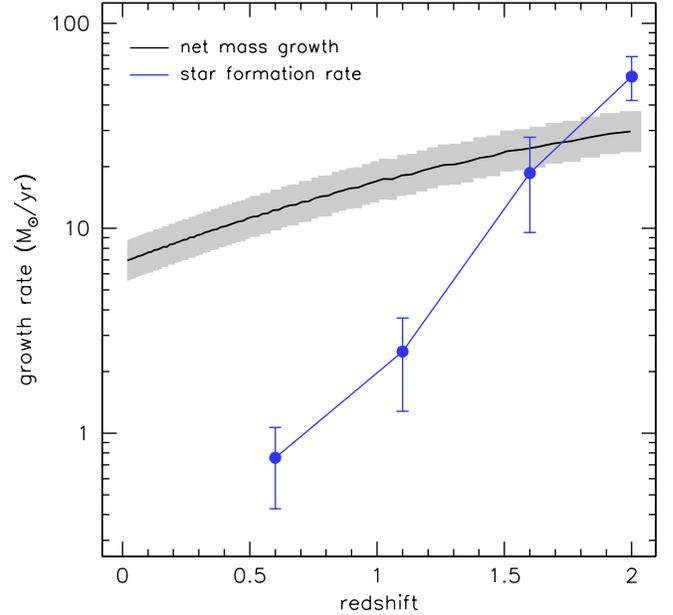}
\caption{\small Growth rate of the galaxies as a function
of redshift, in \msun\,yr$^{-1}$. The net growth rate, derived
from the mass evolution at fixed number density, is indicated
with the black line. The shaded region indicates the $1\sigma$
uncertainty (see \S\,2.2). Blue points with errorbars show the average
star formation rate of the galaxies in each of the stacks, as
derived from fits of stellar population synthesis models to
their SEDs. Star formation can account for most or all
of the observed growth
at $z=1.5-2$, but not for the continued growth at lower redshifts.
\label{sfr.plot}}
\end{figure}

We can assess the contributions of star formation and mergers to the
assembly of the outer parts of massive galaxies as we have
independent measurements of the total mass growth and the growth
due to star formation. The solid line
in Figure \ref{sfr.plot} shows the measured net mass growth (Eq.\ 1)
expressed in \msun\,yr$^{-1}$.
Galaxies with a number
density of $2 \times 10^{-4}$\,Mpc$^{-3}$ have
added mass to their outer regions at a net
rate that declined from $\approx 30$\,\msun\,yr$^{-1}$
at $z=2$ to $\approx 10$\,\msun\,yr$^{-1}$ today.

\noindent
The net mass growth is determined by a combination of mass
growth due to star formation, mass growth due to  mergers,
and mass loss due to winds:
\begin{equation}
\dot{M}_{\rm net} = \dot{M}_{\rm SFR} + \dot{M}_{\rm mergers} -
\dot{M}_{\rm winds}.
\end{equation}
The blue points in Fig.\ \ref{sfr.plot} show the mean star formation
rate $\dot{M}_{\rm SFR}$
of the galaxies that enter each of the stacks. The star formation
rates were 
determined from fits of stellar population synthesis
models to the observed SEDs of the individual galaxies
(see \S\,\ref{nmbs.sec}).
The errorbars were determined from
boostrapping and do not include systematic uncertainties.
As is well known, uncertainties in the star formation histories,
dust content and distribution, the IMF, and other effects can easily
introduce systematic errors of a factor of $\sim 2$ in the star
formation rates, particularly at high redshift
(see, e.g., {Reddy} {et~al.} 2008; {Wuyts} {et~al.} 2009;
{Muzzin} {et~al.} 2009a). The average
star formation rate is similar to the net growth rate at $z=1.5-2$
but significantly smaller at later times. We infer
that the growth of the outer parts of massive galaxies is
not due to a single process but due to a combination of
star formation and mergers. Star formation is only important
at the highest redshifts, and the growth at $z=0-1.5$ is dominated
by mergers.

It is interesting to consider whether the
decline in the star formation rate at $z<1.5$ is directly
related to the structural evolution of the galaxies.
The specific star formation rate of galaxies correlates
well with the average surface density of galaxies within the effective radius,
$\langle \Sigma \rangle
= 0.5 M_{\rm star} / (\pi r_e^2)$, and
there is good evidence for a surface density threshold above which
star formation is very inefficient ({Kauffmann} {et~al.} 2003b, 2006).
Recently {Franx} {et~al.} (2008) have shown that this correlation
exists all the way to $z\sim 3$, and that the threshold evolves with
redshift. The average surface density of galaxies in our study follows
directly from the masses and radii; since $r_e \propto (1+z)^{-1.3}$ and
$M \propto (1+z)^{-0.7}$, we find that $\Sigma \propto (1+z)^{2}$.
Interestingly, the surface densities of our galaxies
are close to the threshold surface density of
{Franx} {et~al.} (2008) and Kauffmann et al.\ (2003b)
above which little or no star formation takes place.
We note that these studies focus on galaxies with lower, more
typical masses than the extreme objects considered here.
Franx et al.\ (2008) noted that the specific star formation
rate may be better correlated with (inferred) velocity dispersion than
with surface density. We later estimate velocity dispersions
for our galaxies, and these do indeed imply little star formation
at $z=0-1$ and increased star formation at $z=2$, if we use
the relation of Franx et al.\ (2008).
We will return to the rapid decline of the star formation rate
in \S\,5.

Quantifying the contributions of star formation and mergers
to the stellar mass at $z=0$
requires an estimate of $\dot{M}_{\rm winds}$, the stellar
mass that is lost to outflows. For a
{Kroupa} (2001) IMF, approximately
50\,\% of the stellar mass that was formed at $z=1.5-2$
was subsequently shed in stellar winds, with most 
of the mass loss occuring in the first 500\,Myr after formation.
It is not clear what happens to this gas. It may
cool and form new stars, still be present in
massive elliptical galaxies in diffuse form (e.g., {Temi}, {Brighenti}, \& {Mathews} 2007),
or lead to a ``puffing up'' of the galaxies if it is removed by
stripping or other effects (e.g., {Fan} {et~al.} 2008).
Irrespective of the fate of this gas, it will not be included in
stellar mass estimates of nearby galaxies, and mass
loss needs to be taken into account when comparing the
integral of the star formation history from $t=t_1$ to $t=t_2$
to the total stellar mass in place at $t=t_2$
(see, e.g., {Wilkins}, {Trentham}, \&  {Hopkins} 2008; {van Dokkum} 2008, and many other studies).

\begin{figure}[htb]
\epsfxsize=8.5cm
\epsffile{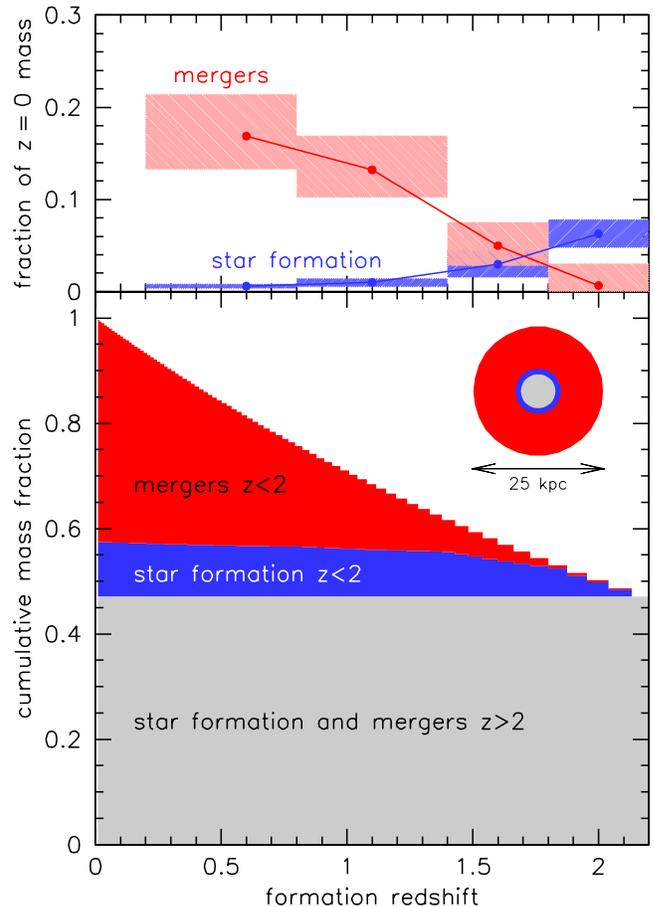}
\caption{\small Contribution of star formation and mergers at $0<z<2$
to the total stellar mass at $z=0$. The top panel shows the contributions
of star formation (blue) and mergers (red) in each of our redshift
bins. To calculate the blue points it was
assumed that 50\,\% of the initial
stellar mass is lost to winds. 
The contributions of mergers were
calculated by subtracting the contributions of star formation from
the total mass growth.
The bottom panel shows the mass build-up over time due to
star formation and mergers. The circles illustrate the mean effective
radius of galaxies at $z=2$ (grey), $1.4<z<2$ (blue; star formation
dominates), and $0<z<1.4$ (red; mergers dominate).
\label{fractions.plot}}
\end{figure}

We calculate the contribution of star formation at $0<z<2$ to the
total mass at $z=0$ by integrating the observed star formation rate over
each redshift interval and applying a 50\,\% correction factor
to account for mass loss. It is assumed that the star formation rate
is constant within each redshift bin. As shown in the top panel
of Fig.\ \ref{fractions.plot} only $6\,\% \pm 2\,\%$ of the total
stellar mass at $z=0$ can be attributed to star formation at $1.8<z<2.2$,
despite  the relatively
high mean star formation rate of galaxies at these redshifts
($55 \pm 13$\,\msun\,yr$^{-1}$). The reason is simply that the time
interval from $z=2.2$ to $z=1.8$ is only 640\,Myr. At lower redshifts
the star formation rate drops rapidly, and the contribution to the
$z=0$ stellar mass declines as well. The bottom panel of
Fig.\ \ref{growth.plot} shows that star formation at $0<z<2$ can
account for only $\sim 10$\,\% of the total stellar mass at $z=0$.

\noindent
The contribution of mergers was calculated by subtracting the
contribution of star formation from the total mass growth.
In the highest redshift bin the contribution of mergers is
very uncertain, but mergers at lower redshift
contribute substantially to the $z=0$ mass.
The growth rate due to mergers is consistent with a roughly constant
value of $\sim 10$\,\msun\,yr$^{-1}$ over the entire redshift
range $0<z<2$. As the mass evolves by a factor of 2 since
$z=2$, the ``specific assembly rate'' (i.e., the growth rate
due to mergers divided by mass) actually
{\em in}creases with redshift by
about a factor of 2. The merger rate can be parameterized as
$dM/M = a (1+z)^m$, and we find $a \sim 0.03$\,Gyr$^{-1}$
and $m\sim 1$ for our sample (see, e.g., {Patton} {et~al.} 2002; {Conselice} {et~al.} 2003, and many other
studies).

As shown in
the bottom panel of Fig.\ \ref{fractions.plot} some
40\,\% of the total stellar mass at $z=0$ was added through
mergers at $0<z<2$.
The circles in the bottom panel of Fig.\ \ref{fractions.plot}
illustrate the increase in the effective radius from $z=2$
to $z=0$. Star formation dominates the growth at $1.5<z<2$
and may be responsible for the increase in $r_e$
over this redshift range. Mergers dominate at lower redshifts
and are plausibly responsible for the size
increase at $0<z<1.5$. 

\subsection{Color Gradients}
\label{colgrad.sec}

If star formation dominates the growth of galaxies at $z=1.5-2$
and this growth mostly occurs at $r\gtrsim r_e$, one might expect
that the galaxies exhibit significant color gradients at these
redshifts. The gradients would be analogous to those in spiral
galaxies, which usually have red bulges composed of old stars
and blue disks with ongoing star formation. We measure color
gradients
by comparing surface brightness profiles of stacks in different
bands. We only use the NMBS near-IR data as it is
difficult to stack the optical CFHT images: the galaxies are
typically very faint in the optical bands, and it is difficult
to fully remove the light from the numerous blue galaxies in
the field. We define a $J^*-K^*$ color, with $J^* = J_1 + J_2$
and $K^* = H_2 + K$. At $z=2$ this color roughly corresponds to
rest-frame $U-R$.

\begin{figure}[htb]
\epsfxsize=8.5cm
\epsffile{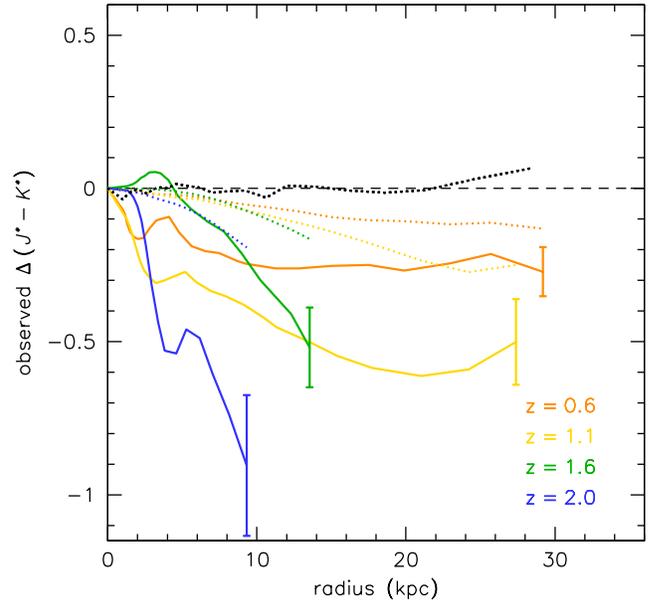}
\caption{\small Radial color profiles
as a function of redshift, in
observed $J_* - K_*$ (see text). The profiles were measured in
the deconvolved stacks (solid lines) and in the original stacks
(broken lines). Typical errorbars, derived from bootstrapping
the stacks, are indicated at the end of each
profile. The black broken line shows the profile of the
stacked stellar image
from $r=0\arcsec$ to $r=3\farcs 6$. The stacked galaxies become bluer
with increasing radius, just like galaxies
at $z=0$. The gradient is large at $z=2$, consistent with
star formation occuring at large radii.
\label{colgrad.plot}}
\end{figure}

Radial color profiles for the deconvolved stacks
are shown in Fig.\ \ref{colgrad.plot} (solid lines).
The data are obviously noisy but show a clear trend: the galaxies are
bluer with increasing radius at all redshifts.
The errorbars are derived from bootstrapping the stacks and do not
include systematic errors due to the deconvolution. Although some
artifact in the deconvolution process may influence the results,
the gradients are robust as the same trends are present in the
original (convolved) stacks (dotted lines).
As expected the stacked stellar image (see \S\,\ref{stacks.sec};
indicated by the black dotted line in Fig.\ \ref{colgrad.plot})
shows no appreciable trend
with radius, demonstrating
that the PSFs are well matched in the different bands.\footnote{The
stellar profile was converted from arcseconds to kpc using the median
conversion factor of the galaxies.}
Although the color gradients are
qualitatively consistent
with the fact that blue galaxies at high redshift are
larger than red galaxies (e.g., {Toft} {et~al.} 2007; {Zirm} {et~al.} 2007; {Franx} {et~al.} 2008),
it is not the same measurement: if the (large) blue and
(small) red galaxies
that enter our stacks had no color gradients we would not measure
a gradient from the stack, as the images in each band are independently
normalized. 

There is an indication that the profiles steepen with redshift, with the
$z=2.0$ stack having the largest color gradient. In the deconvolved stacks
the rest-frame
$U-R$ color at $r>5$\,kpc is 0.5 -- 1 mag bluer than the central
color.
This is a large difference, similar to that
between red sequence and blue cloud galaxies in the
nearby Universe (e.g., {Ball}, {Loveday}, \& {Brunner} 2008). We infer that the
color profiles are
consistent with models in which massive galaxies at $z=1.5-2$
build up stellar mass at large radii through star formation.
The averaged structure of massive galaxies at these redshifts
appears to be qualitatively similar to nearby spiral galaxies, with
a relatively old central component and a young disk. We note,
however, that the galaxies that go into the stacks at these
redshifts have a large range of properties. In particular,
a significant fraction of the population is quiescent and compact
(e.g., {Cimatti} {et~al.} 2008; {van Dokkum} {et~al.} 2008).
A full description
of massive galaxy evolution requires high quality data on large
numbers of individual objects; so far,
such data have only been collected
for small samples (see, e.g., {Genzel} {et~al.} 2006; {Wright} {et~al.} 2009; {Kriek} {et~al.} 2009b).

\noindent
Irrespective of the physical cause of the observed gradients
the immediate consequence is that the galaxies have gradients
in $M/L$ ratio, such that the surface mass density for a given
surface brightness is highest in the center (see {de Jong} 1996; {Bell} \& {de Jong} 2001).
The galaxies are therefore more compact in mass than in light.
This is also the case at low redshift, as elliptical galaxies and
spiral galaxies also have color gradients. However, the effect
may be stronger at higher redshift, which would imply that the
evolution in the mass-weighted effective radius is (even)
stronger than in the luminosity-weighted radius. 
Several
authors have suggested the opposite effect, i.e., that the
sizes of high redshift galaxies may have been underestimated because
of positive gradients in $M/L$ ratio. For example, in the models
of {Hopkins} {et~al.} (2009b) early-type galaxies form in mergers of
spiral galaxies. Owing to star formation in the newly-forming
core merger remnants have blue centers and red outer regions
until $\gtrsim 0.5$\,Gyr after the merger, when the color gradient
starts to reverse.  
{La Barbera} \& {de Carvalho} (2009) take this a step further, as they
infer from color gradients of nearby galaxies that the apparent
size evolution of massive galaxies can be entirely
explained by a constant surface mass density profile combined
with a strong radial age gradient.
As noted above the actual effect is probably the opposite, 
which means that the evolution in Fig.\ \ref{growth.plot}
could be even stronger and the mass in the central 5\,kpc
(Fig.\ \ref{inout.plot}) may actually {\em increase} with redshift.
However,
given the large uncertainties
we did not correct any of our
results for gradients in $M/L$ ratio.

\subsection{Implied Kinematics}
\label{kin.sec}

As noted in many previous studies, high mass galaxies with
relatively small effective radii are expected to have relatively
high velocity dispersions, as the dispersion scales
with $\sqrt{M/r_e}$
(e.g., {Cimatti} {et~al.} 2008; {van Dokkum} {et~al.} 2008; {Franx} {et~al.} 2008; {Bezanson} {et~al.} 2009).
Velocity dispersions at high redshift
provides constraints on the ratio of the
stellar mass to the dynamical mass.
Furthermore, as noted by, e.g.,
{Hopkins} {et~al.} (2009c) and {Cenarro} \& {Trujillo} (2009) the observed
evolution of the velocity dispersion
at fixed stellar mass may help distinguish between physical
models for the size growth of massive galaxies.

It has been possible for some time to measure gas kinematics
of star forming galaxies at high redshift
(e.g., {Pettini} {et~al.} 1998; {Erb} {et~al.} 2003; {F{\"o}rster Schreiber} {et~al.} 2006). The
interpretation is complicated by the fact that the gas disks
are not always relaxed (e.g., {Shapiro} {et~al.} 2008) and by the
fact that  massive
star forming galaxies tend to be systematically larger than
massive quiescent galaxies (e.g., {Toft} {et~al.} 2007; {Zirm} {et~al.} 2007).
Quiescent galaxies generally lack strong emission lines, and
their kinematics can only be measured from stellar absorption lines.
Recently, the first such data
have been obtained. {Cenarro} \& {Trujillo} (2009) and
{Cappellari} {et~al.} (2009) measured velocity dispersions of compact
galaxies at $z=1.5-2$, using deep optical spectroscopy.
{van Dokkum} {et~al.} (2009a) determined the velocity dispersion of a
very small, high mass galaxy at $z=2.2$ from extremely deep
near-IR spectroscopy. From these early results it appears that
the observed dispersions are consistent with
the measured sizes and masses. As an example from our own work,
{van Dokkum} {et~al.} (2008) predicted a velocity dispersion of
$\sigma_{\rm predict}
\sim 525$\,\kms\ for one of the most compact galaxies in their sample,
and subsequently measured a dispersion of
$\sigma_{\rm obs} = 510^{+165}_{-95}$\,\kms\ ({van Dokkum} {et~al.} 2009a).
This also seems to hold at low redshift: {Taylor} {et~al.} (2009) find
that galaxies in SDSS that are
more compact tend to have higher velocity dispersions,
although we note that {Trujillo} {et~al.} (2009) do not see the same
trend in their analysis of SDSS daa.

So far, most studies have considered evolution of the velocity
dispersion at fixed mass, which is obviously not the same as the
actual evolution of the dispersion of any galaxy. Furthermore,
the analysis is usually limited to quiescent galaxies.
As noted by
{Franx} {et~al.} (2008); {Hopkins} {et~al.} (2009c); {Bezanson} {et~al.} (2009); {Cenarro} \& {Trujillo} (2009) and others, a
proper comparison would consider all progenitors, not just the
quiescent galaxies, and explicitly take mass evolution
into account. In the present study we independently measure the
mass evolution and the size evolution at fixed number density,
which allows us to predict the evolution of the velocity dispersion
in a self-consistent way. We calculate the expected dispersion
from the relation
\begin{equation}
M = r_e \langle \sigma^2 \rangle \frac{s_G}{G},
\end{equation}
where $M$ is the total mass, $\langle \sigma \rangle$ is the
average line-of-sight
velocity dispersion over the whole galaxy, weighted by
luminosity, and $s_G$ is the dimensionless gravitational
adius (see, e.g., {Binney} \& {Tremaine} 1987; {Djorgovski} \& {Davis} 1987; {Ciotti} 1991).
As shown by {Ciotti} (1991) the gravitational radius is
a (fairly weak) function of $n$, the Sersic index. A polynomial
fit to the {Ciotti} (1991) numerical results,
\begin{equation}
s_G = 3.316 + 0.026 n - 0.035 n^2 + 0.00172 n^3,
\end{equation}
is accurate to $<0.005$ dex over the range $n=2-10$.

\begin{figure}[t]
\epsfxsize=8.5cm
\epsffile{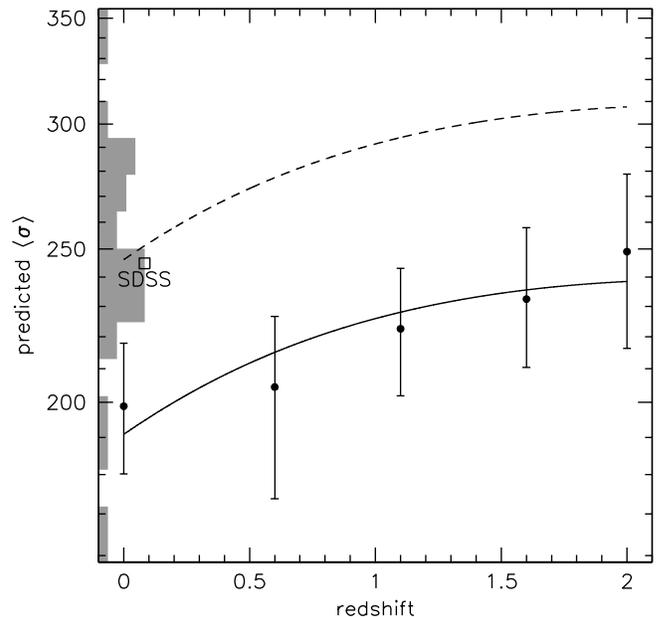}
\caption{\small Expected evolution of the mean luminosity-weighted
velocity dispersion. Points and the
solid line assume that $M_{\rm tot} = M_{\rm star}$, and should
therefore be considered lower limits. The broken line has
the same form as the solid line but is shifted to match the
observed median velocity dispersions of SDSS galaxies (square) and
$z=0$ elliptical galaxies (grey histogram; see Appendix D) with masses
$\log M \sim 11.45$\,\msun. The mean velocity dispersion of galaxies
with a number density of $n=2 \times 10^{-4}$\,Mpc$^{-3}$
is expected to increase with redshift, even though their
masses decrease by a factor of $\sim 2$. Note that the scatter
in $\log \sigma$
is expected to be considerable at each redshift.
\label{sig.plot}}
\end{figure}

The resulting redshift dependence of the
luminosity-weighted line-of-sight velocity dispersion
is shown in Fig.\ \ref{sig.plot}. The points are calculated from
the observed $r_e$, $n$, and stellar mass at each redshift.
The uncertainties are dominated by the uncertainty in the mass
evolution. The solid
line is the evolution that is implied by Eqs.\ \ref{mevo.eq}, \ref{re.eq},
and \ref{n.eq}. The predicted dispersion increases
with redshift by $\approx 0.1$ dex
despite the fact that the masses decrease by a factor of $\approx 2$
over this redshift range. The reason for this counter-intuitive
effect is that the
effective radius decreases more rapidly with redshift
than the mass.

\noindent
The normalization of the curve is uncertain.
The point labeled ``SDSS'' is the median dispersion
of galaxies in SDSS
with a median stellar
mass of $\log M_{\rm star} = 11.45$ in a $\pm 0.15$ dex bin
(obtained from the NYU Value Added Galaxy Catalog; Blanton et al.\ 2005).
The grey histogram shows the measured dispersions of the galaxies
from the OBEY sample ({Tal} {et~al.} 2009) that make up our $z=0$ stack
(see Appendix D). The median dispersion is 245\,\kms, very similar to the
median dispersion of the SDSS galaxies. Note that there is a large range,
with the highest value ($\sigma = 342 \pm 17$\,\kms) measured for NGC\,1399,
the central galaxy in Fornax ({J\o{}rgensen}, {Franx}, \&  {Kj\ae{}rgaard} 1995).\footnote{This galaxy
has a complex dynamical structure in the central regions, as the maximum
dispersion of $\approx 500$\,\kms\ is reached $0\farcs 5$ away from
the center ({Gebhardt} {et~al.} 2007).}
There are no data at higher redshift that can be used, as
to our knowledge no
kinematic studies of samples that are complete in stellar mass
have been done.
The measured $z=0$ dispersions are
offset by $\approx 0.1$ dex from the predictions. This is
not surprising as real galaxies have dark matter, gradients in
$M/L$ ratio, and are not spherical. Furthermore, the SDSS and OBEY dispersions
are measured in a fixed
aperture (or corrected to the value at $r=0$),
and are not identical to the luminosity-weighted mean
dispersion. Scaling the predictions to match the $z=0$ data
leads to a predicted median luminosity-weighted line-of-sight
dispersion of $\sim 300$\,\kms\ at $z=2$.
{Hopkins} {et~al.} (2009c) suggest that the relative
contributions of dark and luminous
matter to the measured kinematics may be a function of redshift,
which could change the evolution in Fig.\ \ref{sig.plot}.
Cold gas may also contribute a non-negligible fraction of the mass
at $z\sim 2$. We have also ignored
the apparent evolution of
color gradients (\S\,\ref{colgrad.sec}): the $z=2$ galaxies
are very blue in the outer parts, and their (mass-weighted)
effective radii are almost certainly significantly overestimated.
Another complication is that the luminosity-weighted average
dispersion is not necessarily the same as the measured dispersion
within an aperture. Interestingly, high redshift data should
be closer to this average than low redshift data as the
aperture is larger in physical units at higher redshift.

Finally, we stress that the evolution in Fig.\ \ref{sig.plot} is
for complete samples of a given (evolving) mass. This
includes star forming galaxies, which probably outnumber
quiescent galaxies at $z=2$
(e.g., {Papovich} {et~al.} 2006; {Kriek} {et~al.} 2006). 
Star forming galaxies are larger than quiescent galaxies
at a given mass and redshift
(e.g., {Trujillo} {et~al.} 2006; {Toft} {et~al.} 2007; {Zirm} {et~al.} 2007; {Franx} {et~al.} 2008; {Williams} {et~al.} 2009),
and we can therefore expect the subset of
quiescent galaxies at $z=2$ to have
dispersions that are significantly larger than indicated
in Fig.\ \ref{sig.plot}. Even within the sample of
quiescent galaxies at $z\sim 2$ the scatter in size
(and hence velocity dispersion) is substantial
(e.g., {Williams} {et~al.} 2009);
as an example, the predicted velocity dispersions of the nine
$z\approx 2.3$ galaxies in
{van Dokkum} {et~al.} (2008) range from $\sim 280$\,\kms\
to $\sim 540$\,\kms. This is of course no different at $z=0$,
as clearly indicated by the grey histogram in Fig.\ \ref{sig.plot}
(see also, e.g., {Djorgovski} \& {Davis} 1987).

\section{Summary and Conclusions}
\label{conclusions.sec}

In this paper we study samples of galaxies at $0<z<2$ with a constant
number density of $2 \times 10^{-4}$\,Mpc$^{-3}$. At low
redshift galaxies
with this number density have a stellar mass
of $3\times 10^{11}$\,\msun\ and
live in halos of mean
mass $\sim 5 \times 10^{13}$\,\msun\ (e.g., {Wake} {et~al.} 2008; {Brown} {et~al.} 2008),
i.e., massive groups. They are mostly the central galaxies in these
groups; only $\sim 10$\,\% are satellites (typically in clusters).
This number-density selection is complementary
to other selection techniques. The main advantage is that it allows
a self-consistent comparison of galaxies at different redshifts,
even if galaxies undergo mergers.
High mass galaxies tend to merge
with lower mass galaxies
(see, e.g., {Maller} {et~al.} 2006; {Brown} {et~al.} 2007; {Guo} \& {White} 2008, and Appendix A),
which means that
their number density remains roughly constant while their mass grows.
The assumption is not that massive galaxies only evolve passively,
but that a large fraction of the most massive galaxies
at $z=0$ had at least one progenitor at higher redshift which was
also among the most massive galaxies.
An important drawback of this selection is that it can
only be usefully applied to galaxies on the exponential tail of
the mass function. A number density selection was previously applied
by {White} {et~al.} (2007), {Brown} {et~al.} (2007, 2008), and
{Cool} {et~al.} (2008) to luminous red galaxies
at $0.2<z<1$.

The stellar
mass of galaxies with a number density $n=2 \times 10^{-4}$\,Mpc$^{-3}$
has evolved by a factor of $\approx 2$ since $z=2$.
To our knowledge this is the first measurement of the mass evolution of
the most massive galaxies over this redshift range. Previous studies
have determined the evolution of the global mass density and the
mass and number density down to fixed mass limits
(e.g., {Dickinson} {et~al.} 2003; {Rudnick} {et~al.} 2003, 2006; {Fontana} {et~al.} 2006; {P{\'e}rez-Gonz{\'a}lez} {et~al.} 2008; {Marchesini} {et~al.} 2009),
but this is a subtly different measurement. On the exponential tail
of the mass function the number density
at fixed mass can change by factors of 5--10 for relatively
small changes in mass. This complicates the interpretation of the evolution
of the mass density, and also makes it highly susceptible to small errors
in the masses (see also {Brown} {et~al.} 2007). Nevertheless,
we note that our results
are consistent with previous studies of the mass function,
and particularly with reports that the high mass end
of the mass function does not show strong evolution
(e.g., {Fontana} {et~al.} 2006; {Scarlata} {et~al.} 2007; {Marchesini} {et~al.} 2009; {Pozzetti} {et~al.} 2009).
At lower redshifts we can
compare our results to other work more directly.
{Brown} {et~al.} (2007) assessed the evolution of the most luminous red galaxies
at $0<z<1$ in a similar way as is done in this study, namely by
determining the evolution of the absolute magnitude of galaxies with
a space density of $4.4 \times 10^{-4}\,$Mpc$^{-3}$
(converted to our cosmology and to
units of dex$^{-1}$ rather than mag$^{-1}$). Their sample selection
does not include blue galaxies, but these are rare in this mass and
redshift range. Using stellar population synthesis models to interpret
the evolution of the absolute magnitude, {Brown} {et~al.} (2007) find that
$\approx 80$\,\% of the stellar mass of the most luminous red galaxies
was already in place at $z=0.7$. This is almost exactly the mass
evolution that we find here: Eq.\ \ref{mevo.eq} implies
that 79\,\% of the mass is in place at $z=0.7$. It is also consistent with 
a later study by
{Cool} {et~al.} (2008) and it is qualitatively
consistent with the evolution of the halo occupation distribution of red
galaxies ({White} {et~al.} 2007; {Wake} {et~al.} 2008).
Despite this consistency with other work systematic errors in the
masses remain the largest cause for concern. As clearly demonstrated
by Muzzin et al.\ (2009a, 2009b) these uncertainties
cannot be addressed by obtaining deeper data or even (low resolution)
continuum spectroscopy, as nearly identical model
SEDs can have very different $M/L$ ratios.

The main result of our paper is that the mass growth of massive galaxies
since $z=2$ is due to a gradual build-up of their outer envelopes.
We find that the mass in the central regions is roughly constant with
redshift, in qualitative agreement with results of {Bezanson} {et~al.} (2009),
{Hopkins} {et~al.} (2009a), and {Naab} {et~al.} (2009).
From our analysis it appears that the well-known
$r^{1/4}$ surface brightness profiles of elliptical galaxies are not
the result of a sudden metamorphosis, like a caterpillar turning
into a butterfly\footnote{Massive galaxies are actually more like
dragonflies than butterflies: dragonflies undergo incomplete
metamorphosis, and are essentially wingless adults
in their nymph stage ---
not unlike the ``wingless'' $z=2$ galaxies. They also share eating
habits: dragonflies are verocious carnivores, and often practise
cannibalism.},
but due to gradual evolution
over the past 10\,Gyr.
We cannot be certain of this due to the limitations of our stacking
technique: the evolution may appear more gradual than it really is if there
is large scatter among the galaxies that enter the stacks. This is
almost certainly the case at $z=2$ (e.g., Toft et al.\ 2007,
Brammer et al.\ 2009).
Figure \ref{growth.plot} goes some way toward addressing a concern raised by
{Hopkins} {et~al.} (2009a), who suggest that observations may have
missed the low surface brightness envelopes of normal elliptical
galaxies at high redshift and that observers may have
erroneously inferred small effective radii for galaxies at $z=1.5-2$.
However, even deeper data at $z=2$ would be valuable to better
constrain the form of the profiles at $r>15$\,kpc.
We note that
{van der Wel} {et~al.} (2008) already showed that surface
brightness biases may exist in
data of low S/N ratio but that they are likely small --- and have
the opposite sign for reasonable light profiles.

A direct consequence of the observed structural evolution is that
massive galaxies do not evolve in a self-similar way. The structure
of galaxies changes as a function of redshift, which means that
the interpretation of
scaling laws such as the fundamental plane (Djorgovski \& Davis 1987;
Dressler et al.\ 1987)
also changes with redshift. This complicates many studies of the
evolution of galaxies, as these usually either explicitly or
implicitly assume self-similarity (e.g., Treu et al.\ 2005,
van der Wel et al.\ 2006,
van Dokkum \& van
der Marel 2007, Toft et al.\ 2007, Franx et al.\ 2008,
Damjanov et al.\ 2009, Cenarro \& Trujillo 2009, van Dokkum et al.\ 2009a,
Cappellari et al.\ 2009, and many other studies). 
Dynamical modeling of spatially resolved internal
kinematics and density profiles can take structural evolution
explicitly into account. Interestingly, although there is no evidence
for departures from simple virial relations in clusters at $z\approx 0.5$
(van der Marel \& van Dokkum 2007), there are indications of such
effects in rotationally supported field galaxies at $z\sim 1$
(van der Wel \& van der Marel 2008).

From the star formation rates of galaxies that enter the stacks
we infer that the physical mechanism that dominates
the build-up of the outer regions  since $z=1.5$ is likely some form
of merging or accretion, consistent with
many previous studies
(e.g., {van Dokkum} {et~al.} 1999; {van Dokkum} 2005; {Tran} {et~al.} 2005; {Bell} {et~al.} 2006; {White} {et~al.} 2007; {McIntosh} {et~al.} 2008;
Naab et al.\ 2007, 2009).
In-situ star formation may dominate
the growth at $z=1.5-2$, but the newly formed stars
account for only $\sim 10$\,\% of the total stellar
mass at $z=0$ --- about 1/4 of the contribution of mergers.
The distinction between star formation and mergers
is obviously somewhat diffuse at high redshift,
as star forming disks may be continuously replenished
(see, e.g., {Genzel} {et~al.} 2008; {Franx} {et~al.} 2008; {Dekel} {et~al.} 2009). Furthermore, the galaxies that are accreted at $0<z<1$ may
well have formed some fraction of their stars at $1<z<2$.
It seems likely that star formation also dominated at $z>2$;
as noted by many authors, the formation of the compact cores
of elliptical galaxies was almost certainly a highly dissipative
process (see, e.g., {Kormendy} {et~al.} 2009, and reference{}s therein).
It is unknown why star formation shuts off at later times;
this could be due to feedback from an active nucleus
(e.g., {Croton} {et~al.} 2006; {Bower} {et~al.} 2006), virial shock heating of the gas
(e.g., {Dekel} \& {Birnboim} 2006), gravitational heating due to accretion
of gas or galaxies (e.g., {Naab} {et~al.} 2007; {Dekel} \& {Birnboim} 2008; {Johansson}, {Naab}, \&  {Ostriker} 2009),
starvation ({Cowie} \& {Barger} 2008),
or other processes. Interestingly, we find that the shut-off is a
rather sudden event, with the star formation rate dropping by
a factor of 20 from $z=2$ to $z=1.1$ whereas the stellar mass grows only
by a factor of 1.4 over this redshift range. This may suggest that the
quenching trigger is not only a simple (stellar) mass threshold, as the
range of masses in our selection bin is a factor of 2 at each
redshift --- larger than the evolution in the median mass.
We note that the stellar mass threshold that we would derive is
$\approx 2 \times 10^{11}$\,\msun.
Another open question is what the star formation histories
are of the galaxies that are accreted (see, e.g., Naab et al.\ 2009).
The properties of the stellar populations of elliptical galaxies
at $r\gg r_e$ can give interesting constraints in this context
(see, e.g., {Weijmans} {et~al.} 2009).

The analysis in this paper can be improved and extended in many ways.
The most obvious is to study the profiles of individual galaxies to
large radii. Even though the stacking procedure should give
reasonably accurate mean radii, the measured mean
profile shape (parameterized
by the Sersic $n$ parameter) can be in error (see Appendix B).
Furthermore, valuable information is obviously lost --- for example,
the rich diversity of massive galaxies at $z\gtrsim 2$ (see Kriek et al.\
2009b) --- and the interpretation
rests on several simplifying assumptions. 
The most important of these may be that all the galaxies that
enter the stacks evolve in a somewhat homogenous way. It
may well be that the samples consist of quite distinct populations
whose relative number fractions change with time. We would interpret
this as smooth evolution, whereas in reality there might be
few individual
galaxies that actually have the mean properties that we measure.
Such effects are likely important 
at $z=1.5-2$ as our sample contains both quiescent and star
forming galaxies at these redshifts, and they form quite distinct
populations (e.g., Kriek et al.\ 2009b, Brammer et al.\ 2009).
The population is likely more homogeneous at lower redshifts.
At present studying surface brightness profiles of individual
galaxies to very faint
limits is only possible at low redshift
(e.g., {Kormendy} {et~al.} 2009, Tal et al.\ 2009),
but progress can be expected from
ongoing deep ground- and space-based surveys. We also assume that
our samples are complete and unbiased at all redshifts, but
there could be biases against very extended galaxies at the
highest redshifts. We verified that individual
galaxies with the properties of the $z=0.6$ stack would be
detected (with approximately the correct flux) at $z=2$, but
more extreme objects may have escaped detection.
It will also be worthwhile
to stack images with better spatial resolution. The highest redshift
galaxies in our study are not resolved within the effective radius,
and this may lead to biases in the Sersic fits.

One of the main uncertainties is the analysis is the conversion
from rest-frame $R$ band light to mass. We know that the $M/L$
ratio is not constant with radius even at $z=0$, and we find
good evidence for strong radial trends at higher
redshift. It seems therefore possible that we might be
overestimating the half-mass radii of galaxies at $z=2$ by a larger
factor than we are overestimating the radii at $z=0$.
We certainly do not see evidence for an
an {\em increasing} $M/L$
ratio with radius, such as predicted by, among others,
{La Barbera} \& {de Carvalho} (2009). Upcoming surveys with WFC3 on HST will resolve
this issue, and allow derivation of mass-weighted radii.
Finally, we have mostly ignored the effects of dark matter in
this paper, and of possible evolution in the IMF (e.g., van Dokkum 2008,
Dav\'e 2008, Wilkins et al.\ 2008). Kinematic data will
give independent information on the masses of galaxies
at high redshift, although it will be difficult to disentangle
the effects of errors in stellar masses,
changes in $s_G$, evolution in the
stellar IMF, and the effects of dark matter.
It will also be interesting to connect the evolution of
these galaxies to the evolution of their halos, by combining
the evolving stellar mass at fixed number density with clustering
measurements and HOD modeling (see, e.g., {White} {et~al.} 2007; {Wake} {et~al.} 2008; {Quadri} {et~al.} 2008).

\acknowledgements{
We thank the referee for a very constructive report, which
improved the paper. Ron Probst and the NEWFIRM team are thanked
for their work on the instrument and help during the observations.
We used the HyperLeda database (http://leda.univ-lyon1.fr).
This paper is partly
based on observations obtained with MegaPrime/MegaCam, a joint project
of CFHT and CEA/DAPNIA, at the Canada-France-Hawaii Telescope (CFHT)
which is operated by the National Research Council (NRC) of Canada,
the Institut National des Science de l'Univers of the Centre National
de la Recherche Scientifique (CNRS) of France, and the University of
Hawaii. We thank the CARS team for providing us with their
reduced CFHT mosaics.
Support from NSF grants AST-0449678 and AST-0807974
is gratefully acknowledged. Part of the work described in this
paper was done during an extended visit of the PI to Leiden University.
}


\newpage
\begin{appendix}
\section{Appendix A. Effects of Mergers on the Mass Function}
\begin{figure*}[htb]
\epsfxsize=16cm
\epsffile[28 540 534 689]{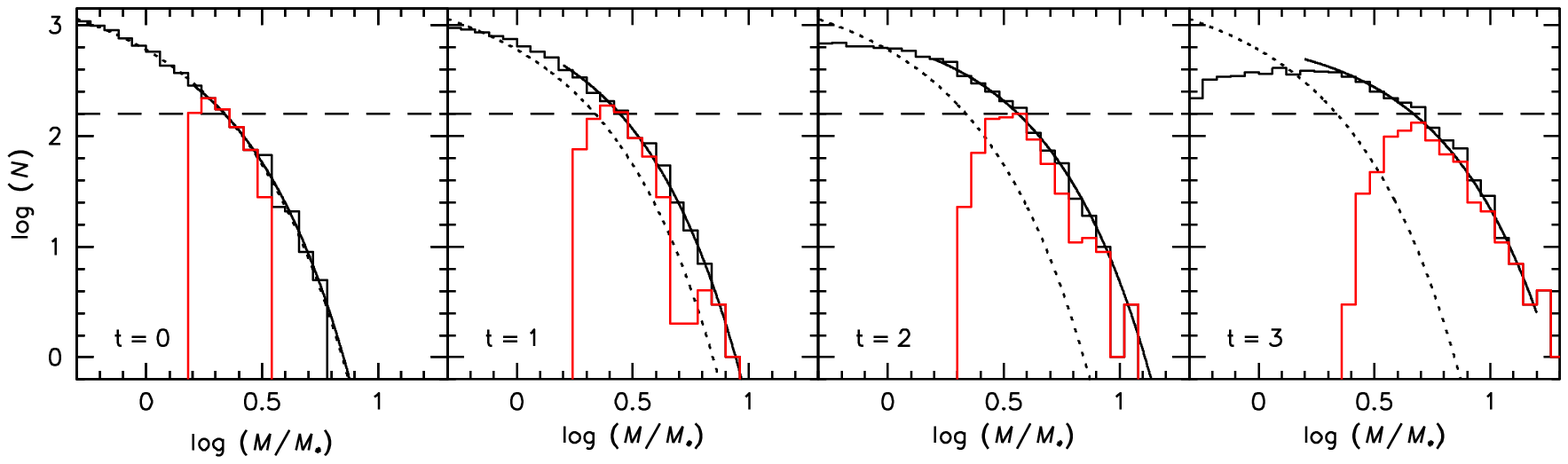}
\caption{\small Monte-Carlo simulation demonstrating the effects of
merging on our constant-number density selection. At $t=0$ galaxies
are distributed according to a Schechter (1976) function (dotted
line in each panel). At each timestep, all galaxies merge with one
other galaxy, reducing the total number density by a factor of 2.
The mass ratio of the mergers is randomly chosen between $1:10$ and
$1:2$. The dashed horizontal line is a line of
constant number density. The red histograms show galaxies with
initial masses of $0.2\leq \log (M/M_*)\leq 0.5$ and their descendants.
Because of the steepness of the mass function at $M>M_*$ the
descendants have roughly the same number density as their progenitors.
\label{sim.plot}}
\end{figure*}

In this paper we select galaxies at a constant
number density, as opposed to a constant luminosity or mass.
At each redshift galaxies are selected in a
narrow mass bin whose median
mass corresponds to the mass appropriate for the chosen number density
(see \S\,\ref{select.sec}). This selection is appropriate for
processes that change the masses of galaxies and not their number
densities, such as star formation and mass loss. However, mergers
change both the mass and the number density of galaxies, and might be
expected to complicate the selection.
The effect of mergers on our selection was assessed with Monte
Carlo simulations. A sample of 50,000 galaxies was created, distributed
according to a {Schechter} (1976) mass function with characteristic
mass $M_* \equiv 1$ and faint-end slope $\alpha = -1.2$, over the
mass range $-1.5 \leq \log (M/M_*) \leq 1.0$. The simulation is
independent of the precise value of $\alpha$; the value we chose
is intermediate between two recent studies
({Marchesini} {et~al.} 2009; {Kajisawa} {et~al.} 2009).
The left-most
panel of Fig.\ \ref{sim.plot} shows the high-mass end of this mass
function. The horizontal line in this panel is an (arbitrarily
chosen) constant number density of $n_{\rm sel} = 32$
galaxies per bin. The red
histogram shows galaxies with masses in the range $0.2 \leq
\log (M/M_*) \leq 0.5$.

\noindent
From time step $t=0$ to $t=1$ all galaxies are merged with each other,
reducing the total number of galaxies by a factor of two.
Mergers of galaxies with
mass ratios between $1:10$ and $1:2$ have equal probability.
The results are qualitatively similar if other limits are assumed, such
as a constant $1:4$ ratio. The red histogram at $t=1$ shows the distribution
of galaxies that have at least one progenitor whose original mass
was in the range
$0.2 \leq \log (M/M_*) \leq 0.5$. The distribution is shifted and
has broadened, but the median mass is very similar to the
mass of galaxies with a number density of $n_{\rm sel}$. Similarly,
the merger remnants are merged with each other from $t=1$ to $t=2$
and again from $t=2$ to $t=3$.

\begin{figure*}[htb]
\epsfxsize=16cm
\epsffile[28 530 534 689]{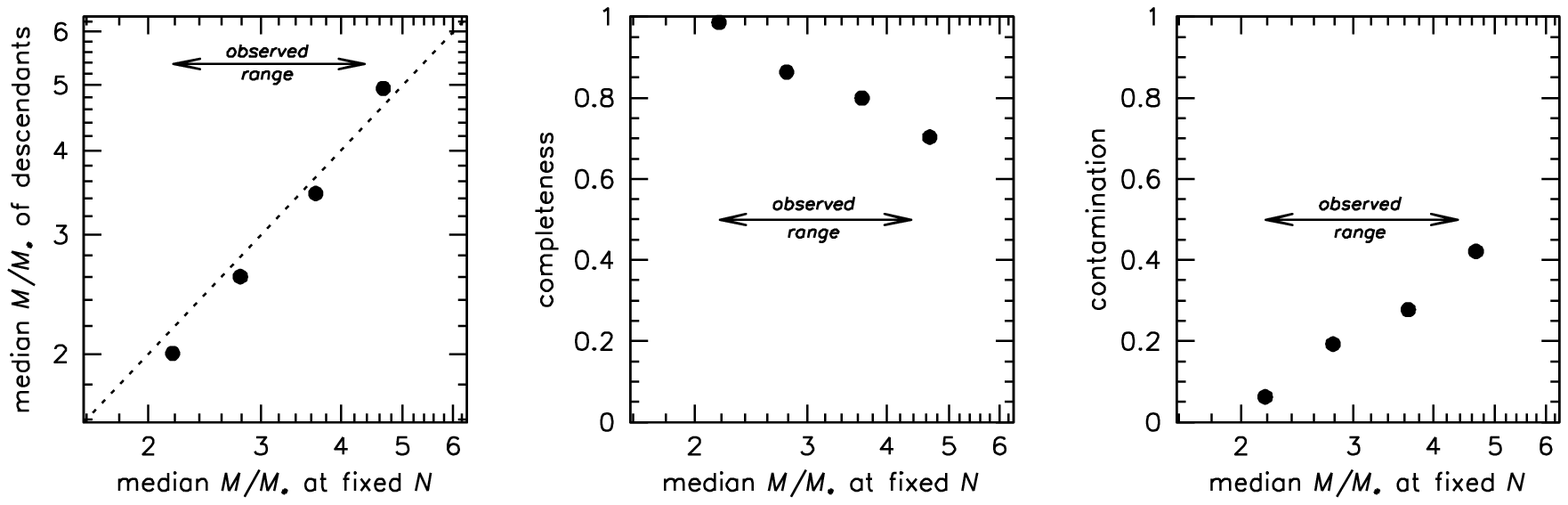}
\caption{\small Results from the Monte Carlo simulation. {\em Left
panel:} Comparison of the actual median mass of the descendants of
galaxies with initial masses $0.2 \leq \log (M/M_*) \leq 0.5$ to
the measured mass at fixed number density (which is used in this paper).
There is excellent agreement.
The double arrow indicates the range of masses measured
at $0<z<2$ (see \S\,\ref{select.sec}). {\em Middle panel:} Fraction of
descendants that is selected with the method described in
\S\,\ref{select.sec}. {\em Right panel:} Fraction of galaxies that
are selected but are not descendants. Over the relevant mass range
the completeness is high and the contamination low.
\label{simresult.plot}}
\end{figure*}

We applied the same selection method as used in \S\,\ref{select.sec} to
the simulated sample. Exponential functions were fit to the high mass
end of the mass function (solid lines in Fig.\ \ref{sim.plot}), and
the intersections of these lines with $n_{\rm sel}$ (horizontal
dashed lines) were determined. The left panel of Fig.\ \ref{simresult.plot}
compares the actual median masses of all descendants of galaxies
with $0.2 \leq \log (M/M_*) \leq 0.5$ at $t=0$ to the measured
masses at fixed number density. There is excellent agreement, demonstrating
that our selection method gives the correct mass evolution for a merging
population of galaxies. Next, we selected galaxies in a mass bin of
width $\pm 0.15$ dex centered on the evolving mass. These bins miss
some of the descendants as their mass distribution broadens with time.
The middle panel of Fig.\ \ref{simresult.plot} shows the fraction of
actual descendants that are contained within the selection bin. This
fraction is $\sim 70$\,\% for mass evolution of
a factor of two. Finally, the right panel shows the fraction of galaxies
in the bin that are not descendants of galaxies with original masses
$0.2 \leq \log(M/M_*) \leq 0.5$. This fraction is $\sim 40$\,\% for
a factor of two mass evolution, but we note that all
of these galaxies had original masses close to the selected range.

\noindent
In summary, a selection at constant number density should give a
fairly homogeneous sample of galaxies as a function of redshift.
Mass evolution can be measured directly in a self-consistent way,
regardless of the physical process (star formation or mergers).
In reality
mergers likely dominate at the high-mass end of the mass function
and star formation likely dominates at the low-mass end
(e.g., {Guo} \& {White} 2008; {Damen} {et~al.} 2009).
We note that our simple simulation
demonstrates that the observed average mass growth of a factor of two
at high masses can
be explained by three mergers with random mass ratio between
$1:10$ and $1:2$.
The selection does not produce homogeneous samples
if growth occurs through $1:1$ mergers only
(and no other mergers), but that is exceedingly unlikely
(see, e.g., {van Dokkum} 2005; {Naab} {et~al.} 2007; {Guo} \& {White} 2008).

\section{Appendix B. Recovering Average Structural Parameters from Stacked Images}

A key aspect of this study is that structural parameters are not
measured from individual images and then averaged but measured from
averaged images of many individual galaxies. We tested how well
the structural parameters $r_e$ and $n$ can be recovered with this
method by creating stacks of simulated galaxies and by analyzing
real galaxies.

\subsection{Model galaxies}

Model galaxies were created with GALFIT ({Peng} {et~al.} 2002). Their Sersic
indices are distributed randomly between $n=1$ and $n=4$, their
axis ratios range from $b/a = 0.1$ to $b/a=1.0$, and they have
random position angles.
The results are not sensitive to the assumed distribution
of $n$ or $b/a$.
Ten stacks were created of 200 galaxies each.
The ten stacks differ in their distributions of circularized
effective radii.
The radii were chosen randomly within the range $0.5 r_n < r_e < 2 r_n$,
with $r_n$ ranging from $r_1=1$\,kpc to $r_{10}=10$\,kpc.
Prior to stacking
the galaxies were placed at $z=2$, convolved with a Moffat PSF
with a FWHM of $1\farcs 1$, and sampled with $0\farcs 3$ pixels.

\begin{figure*}[htb]
\epsfxsize=16cm
\epsffile[30 422 490 640]{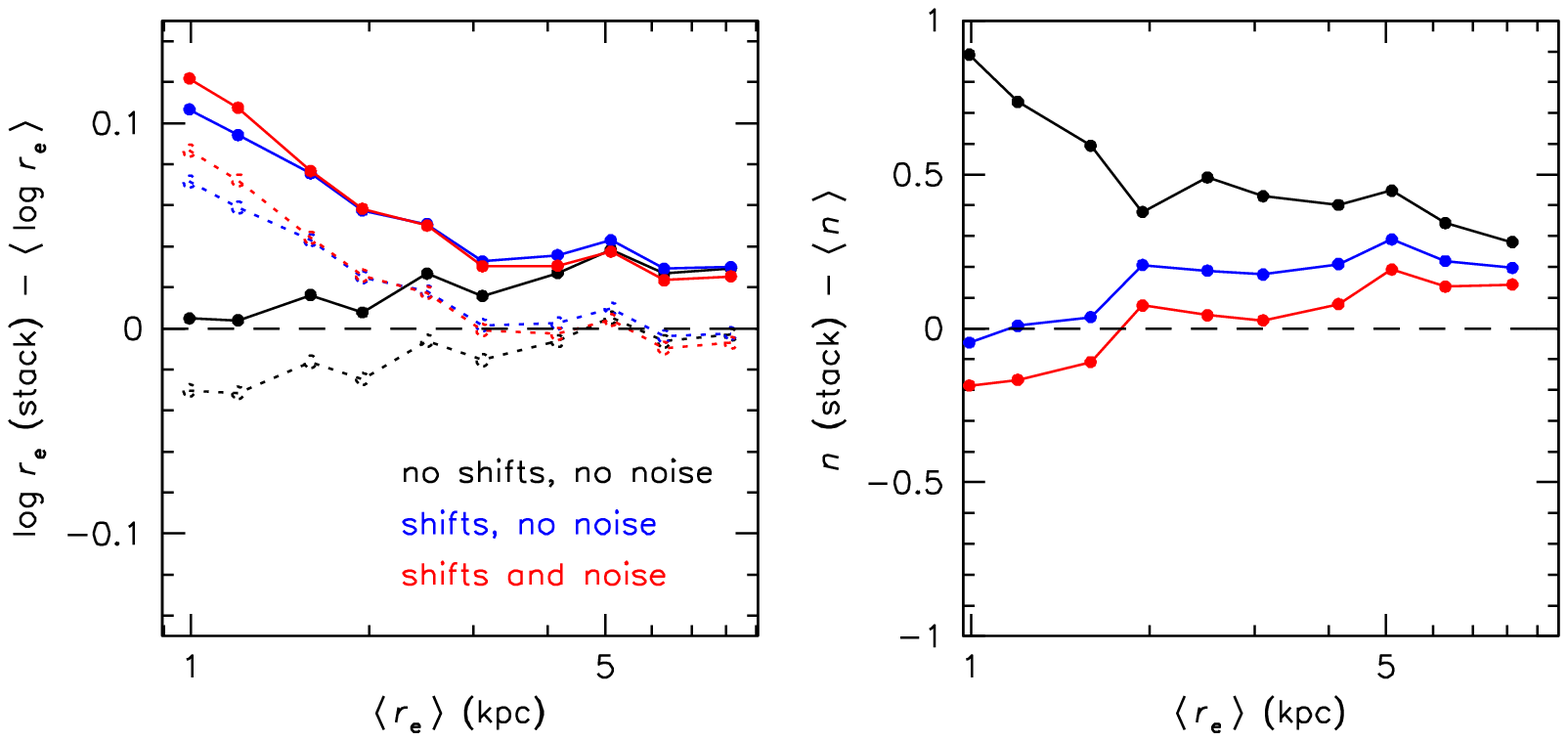}
\caption{\small {\em Left panel:} Comparison of the effective radius
as measured from a stack of 200 galaxies to the mean circularized
effective radius of the individual galaxies. Black is for a noiseless
stack, blue is for a noiseless stack with Gaussian 0.25 pixel shifts
applied to the individual galaxies, and red is for a stack with shifts
and the same noise as the $z=2$ stack of real galaxies. Solid lines show
results for
$\langle \log r_e \rangle$ and broken lines are for
$\log \langle r_e \rangle$. Sizes $\gtrsim 2$\,kpc can
be measured fairly reliably, with a systematic
error of $\sim 10$\,\%. {\em Right panel:} Comparison of
Sersic (1968) index $n$. The stacks tend to overestimate the
true average Sersic index, by $\lesssim 0.5$. Note that stacks
with simulated shifts and noise perform better than noiseless
stacks.
\label{stackcheck.plot}}
\end{figure*}

The stacked images were fit with GALFIT and the results are shown
in Fig.\ \ref{stackcheck.plot} (black lines). The circularized
effective radii are recovered well, even for $\langle r_e \rangle
= 1-2$\,kpc. This is remarkable as these scales
correspond to $0 \farcs 1 - 0 \farcs 2$, a small fraction of
the FWHM of the PSF. Broken lines and solid lines are for two
different ways of averaging the effective radii of individual
galaxies: solid lines show averages of $\log r_e$ and
broken lines are for the logarithm of the average $r_e$. The
stacks clearly measure the average of $r_e$
rather than the average of $\log r_e$ but the differences
are small. The right panel shows how well the average Sersic
index is recovered. The stacks systematically overestimate the
Sersic index. This can be understood by considering the average
profile of a small galaxy and a large galaxy, both with $n=1$.
The small galaxy will add a peak at small radii to the extended
profile of the large galaxy, creating a profile best fit by
a model with larger $n$.

The noiseless test is useful as it demonstrates that the stacking
technique can give results that can be compared directly with
measurements of individual galaxies. However, in order to assess
the systematic errors associated with our methodology centroiding
errors and noise need to be taken into account. Centroiding errors
were simulated by shifting the individual images by small amounts,
using the same third order polynomial
interpolation as was used for the real data. The shifts were
drawn from a Gaussian distribution with $\sigma = 0.25$ pixels
(2.1 kpc). Blue curves in Fig.\ \ref{stackcheck.plot} show the
effects of these shifts on the recovered parameters. As expected,
the smoothing leads to an overestimate of the effective radius.
However, the effect is fairly small because
the profiles at larger radii are not strongly affected by the
centroiding errors. The average Sersic parameter is actually {\em better}
determined when small shifts are included, probably because the
central peak (caused by galaxies with small $r_e$) is smeared out in
the stacks.

Finally, noise was added to the modeled stacks. A noise image was
created from the actual residual map of the $z=2$ stack, thus ensuring
that the noise level, correlations between pixels,
and non-Gaussian components are all exactly identical to the actual
data. The $z=2$ stack has the highest noise of our four stacks
as the galaxies are fainter than those at lower redshift. The noise
images were added to the artificial stacks, and structural
parameters were remeasured. The red curves in Fig.\ \ref{stackcheck.plot}
show the results. They are quite similar to the blue curves,
suggesting that systematic effects dominate over the effects of
noise. In summary, we should be able to determine effective radii and
Sersic $n$ parameters with reasonable accuracy despite the poor
spatial resolution of our data, if the surface brightness
profiles are well described by Sersic fits.

\begin{figure*}[htb]
\epsfxsize=8cm
\begin{center}
\epsffile[65 189 524 645]{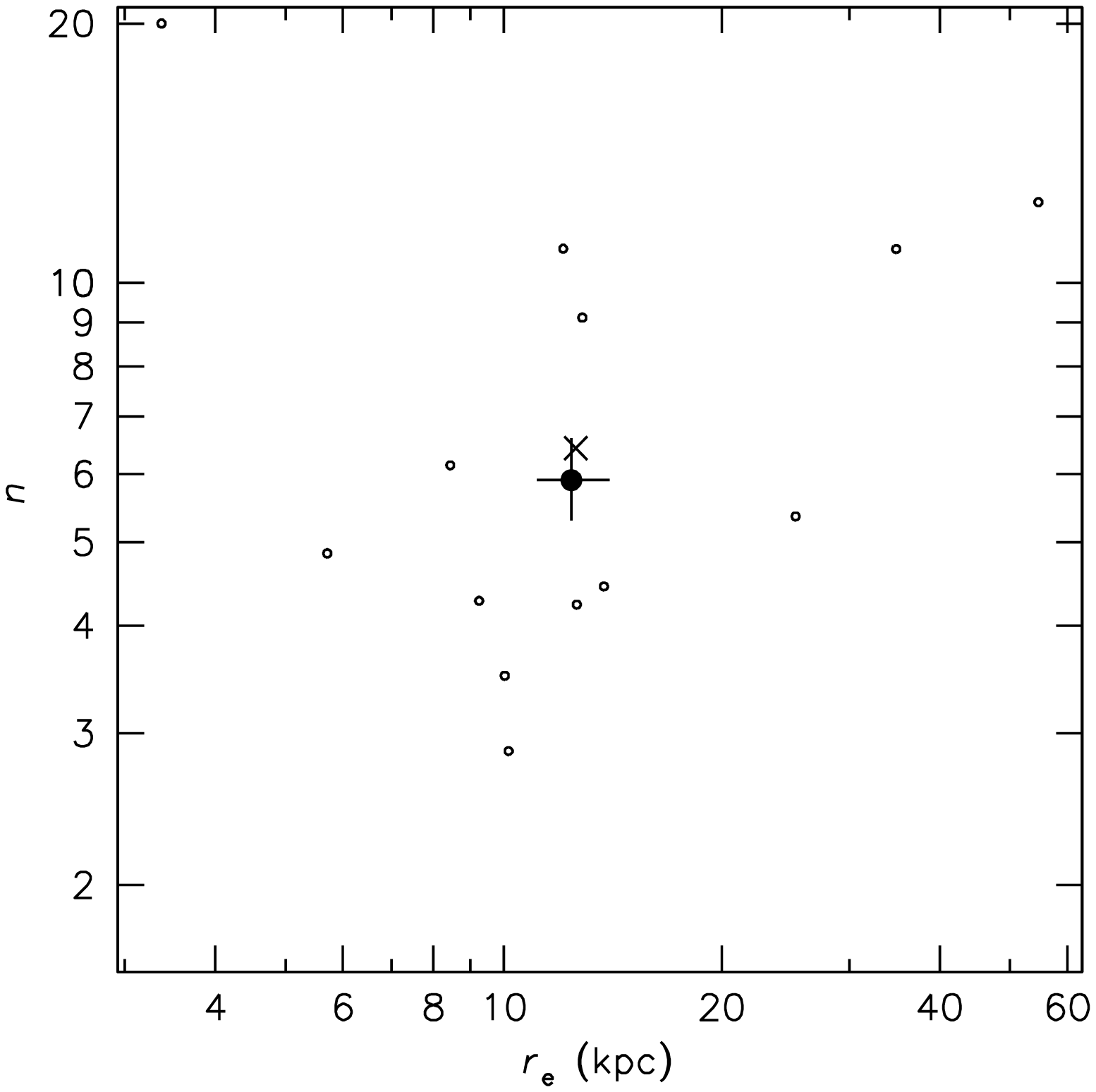}\vspace{-0.5cm}
\end{center}
\caption{\small Comparison of the effective radius and Sersic $n$ parameters
of individual galaxies in the $z=0$ OBEY
sample to the average as measured from the stacked image. The
results from the stack are indicated by the solid circle with errorbars.
The actual means of the individual galaxies are indicated with the
cross. The measurements from the stack are fully consistent with the
means of the individual galaxies.
\label{obeystack.plot}}
\end{figure*}
\begin{figure*}[t]
\epsfxsize=10cm
\begin{center}
\epsffile{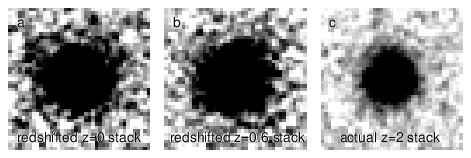}
\end{center}
\caption{\small {\em a)} Stack of $z=0$ OBEY galaxies redshifted
to $z=2$; {\em b)} stack of $z=0.6$ NMBS galaxies
redshifted to $z=2$; and {\em c)} stack of actual NMBS $z=2$
galaxies. The images span $12''\times 12''$, and are normalized
by the peak flux. The contrast is scaled to
bring out differences in the outer regions.
The redshifted low redshift stacks are more extended than the
$z=2$ stack, showing low surface brightness emission out to
large radii.
\label{z2shift.plot}}
\end{figure*}

\subsection{Real galaxies}

Real galaxies do not necessarily follow Sersic profiles, and subtle
deviations for individual galaxies may lead to significant systematic
differences when determining structural parameters from stacked data.
We first test whether the structural parameters that we derive
for the stacked $z=0$ OBEY sample (see Appendix D) are consistent with
the average of the individual galaxies. We fitted Sersic (1968) profiles
and determined circularized
effective radii in kpc and the Sersic $n$ parameter
for each of the 14 galaxies that enter the OBEY stack. The results are
shown in Fig.\ \ref{obeystack.plot}. We find $\langle r_e\rangle
= 12.6$\,kpc and $\langle n \rangle = 6.4$. The structural parameters
measured from the stacked image are very similar
at $r_e = 12.4^{+1.6}_{-1.3}$\,kpc,
$n=5.9^{+0.7}_{-0.6}$, and we conclude that the stacking method
gives reasonable results for real galaxies.

Next, we assess the importance of redshift-dependent effects by
redshifting the $z=0$ OBEY stack and the $z=0.6$ NMBS stack to $z=2$.
The angular scale and fluxes of the profiles of the OBEY galaxies were
corrected to $z=2$, the galaxies were convolved with the NMBS PSF, the
images were resampled to match the NMBS resolution of $0\farcs 3$
pixel$^{-1}$, and empirical noise derived from empty regions of the
actual NMBS images was added. The $z=0.6$ images were only scaled in
flux, as they have  a similar PSF
and spatial scale as the $z=2$ images.
Noise was added in the same way as for the OBEY stack.

The resulting redshifted images are shown in Fig.\ \ref{z2shift.plot},
along with the actual $z=2$ stack. To highlight the differences in
profile shape the images were normalized to the peak flux. It is clear
that the actual $z=2$ image is more compact than the redshifted $z=0$
and $z=0.6$ stacks, as it lacks the low surface brightness features
that surround the bright cores in the lower redshift stacks.  We
quantified the effects of redshifting by fitting Sersic profiles to
the redshifted stacks. The redshifted OBEY stack gives $r_e =
11.0$\,kpc and $n=4.9$, in good agreement with the original values
($r_e = 12.4^{+1.6}_{-1.3}$\,kpc, $n=5.9^{+0.7}_{-0.6}$). The
redshifted $z=0.6$ stack gives $r_e=8.0$\,kpc and $n=4.4$, again in
good agreement with the original values
($r_e=8.0^{+1.2}_{-0.5}$\,kpc, $n=4.0^{+0.4}_{-0.4}$).  From these
tests we conclude that there are no obvious redshift-dependent effects
which could lead to severe biases in the derived evolution.

\section{Appendix C. Individual Galaxy Images}

Here we show the individual images of galaxies that enter the stacks.

\begin{figure*}[htb]
\epsfxsize=16cm
\epsffile{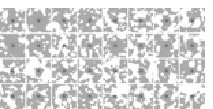}
\caption{The 32 galaxies with $\log M \approx 11.36$ and
$0.2<z<0.8$ that enter the $\langle z\rangle =0.6$ stack. The images
are averages of \Ja\ and \Jb. Masked regions are white. Each image was normalized
using the summed flux in the central $10 \times 10$ pixels ($3\farcs 0 \times
3\farcs 0$); this is why the background noise is not the same for all galaxies.
{\bf \em Note: Figures C1-C4 were degraded to adhere to arXiv's file size
limits.}
\label{mos_0.6.plot}}
\end{figure*}

\begin{figure*}[p]
\epsfxsize=16cm
\epsffile{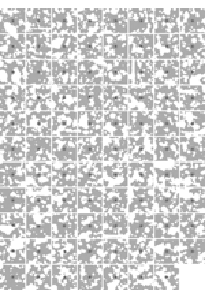}
\caption{\small The 87 galaxies with $\log M \approx 11.28$ and
$0.8<z<1.4$ that enter the $\langle z\rangle =1.1$ stack. The images
are averages of \Jc\ and \Ha.
\label{mos_1.1.plot}}
\end{figure*}

\begin{figure*}[p]
\epsfxsize=16cm
\epsffile{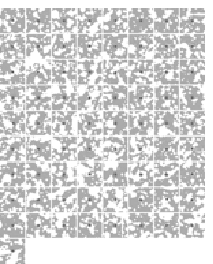}
\caption{\small The 73 galaxies with $\log M \approx 11.21$ and
$1.4<z<1.8$ that enter the $\langle z\rangle =1.6$ stack. The images
are averages of \Ha\ and \Hb.
\label{mos_1.6.plot}}
\end{figure*}

\begin{figure*}[p]
\epsfxsize=16cm
\epsffile{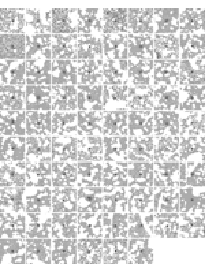}
\caption{\small The 78 galaxies with $\log M \approx 11.15$ and
$1.8<z<2.2$ that enter the $\langle z\rangle =2.0$ stack. The images
are averages of \Hb\ and $K$.
\label{mos_2.0.plot}}
\end{figure*}

\null
\newpage

\section{Appendix D. Low Redshift Galaxies}

\subsection{The OBEY Sample}

The data over the redshift range $0.6<z<2.0$ are analyzed in a self-consistent way, and are all
drawn from the same survey (the NEWFIRM Medium Band Survey, or NMBS). Although essentially
all the results presented in this paper could be derived from the NMBS data alone, we made
some effort to construct a $z=0$ sample that can be analyzed in the same way as the data
at higher redshift. Key requirements are that the masses are on the same system as the
high $z$ data and that very deep photometry is available to probe the faint outer regions
of the galaxies. We use data from a recent public survey of luminous elliptical galaxies,
called Observations of Bright Ellipticals at Yale (OBEY) ({Tal} {et~al.} 2009).
The OBEY sample consists of all elliptical galaxies from the {Tully} (1988) Nearby Galaxies
Catalog with distances\footnote{Distance-dependent quantities refer to the Tully (1988) catalog,
and are for $H_0=75$\,\kms\,Mpc$^{-1}$.} 15 -- 50 Mpc, luminosities $M_B<-20$,
declinations between $-85^{\circ}$ and $+10^{\circ}$, and Galactic latitude $>17^{\circ}$.
The galaxies were observed with the CTIO 1\,m telescope, as described in {Tal} {et~al.} (2009).
Owing to very careful flatfielding the surface brightness profiles can be reliably
traced to large radii. The data are publicly available.\footnote{See www.astro.yale.edu/obey.}

We determined stellar masses for the galaxies in the OBEY sample in the following way.
Total magnitudes and colors were obtained from {Prugniel} \& {Heraudeau} (1998) through the
HyperLeda interface ({Paturel} {et~al.} 2003). The ``extrapolated'' total $B$ magnitudes were used
together with ``effective''
luminosity-weighted $U-B$, $B-V$, $V-R$, and $V-I$ colors to create $UBVRI$ SEDs.
The apparent magnitudes were corrected for Galactic extinction using the estimates
from {Schlegel}, {Finkbeiner}, \&  {Davis} (1998) and converted to
absolute magnitudes using the distances given in {Tully} (1988) (corrected to
our cosmology).
Stellar masses were determined using FAST
({Kriek} {et~al.} 2009a), a code that fits stellar population synthesis models to observed
photometry. This code was also used to determine the stellar masses of the galaxies in
the NMBS, and we used the same stellar population synthesis model, dust law, and other
parameters as were used for the fits to the distant galaxies (see \S\,\ref{nmbs.sec}).
The only difference is
that we fixed the value of the characteristic star formation timescale $\tau$ to
1\,Gyr. If $\tau$, age, and $A_V$ are all free parameters the fits show aliasing
due to the limited number of data points. We verified that this small change to
the fitting procedure does not lead to systematic biases in the masses.

\begin{figure*}[htb]
\epsfxsize=11cm
\epsffile[-130 206 350 556]{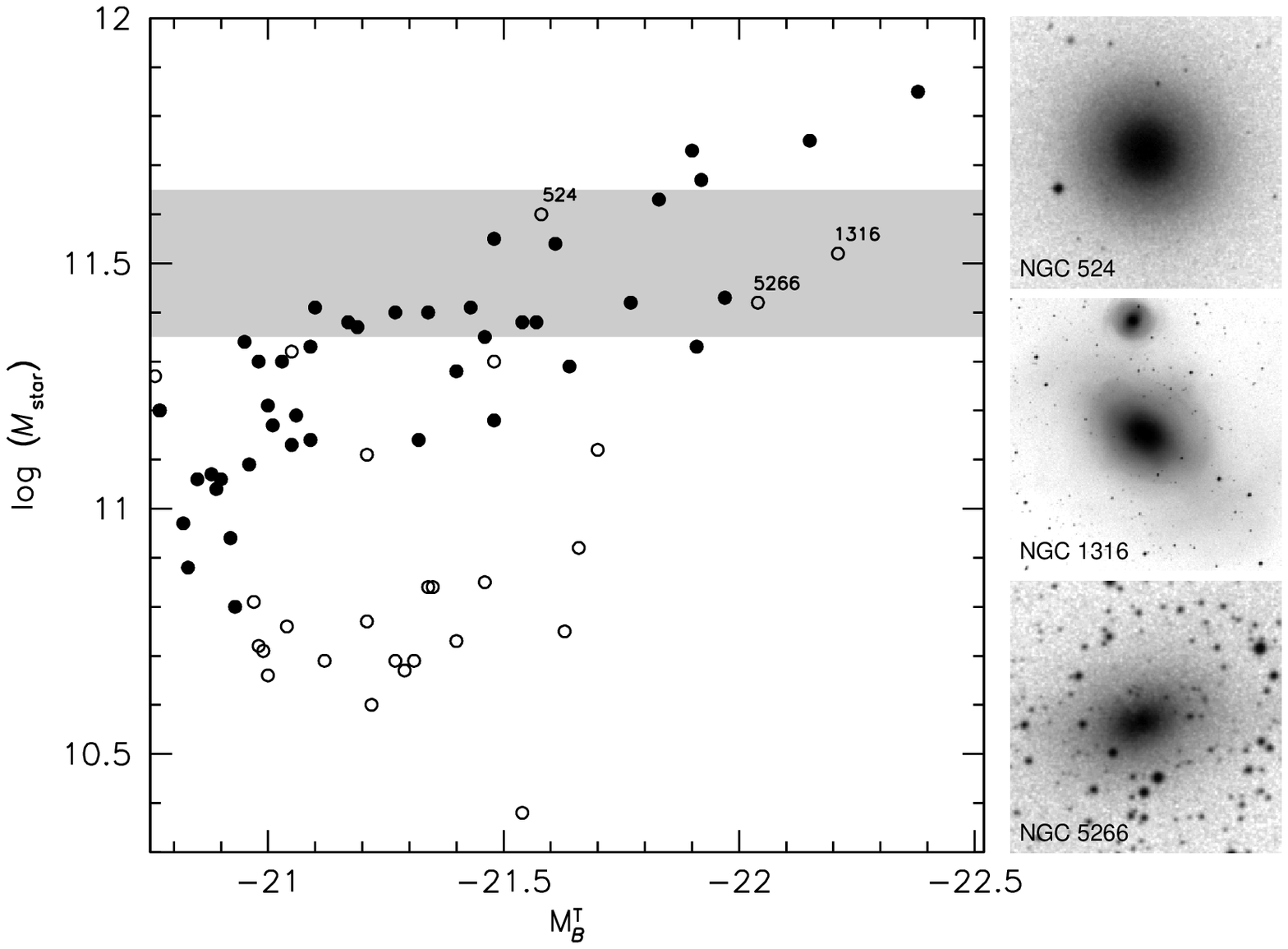}
\caption{\small Relation between stellar mass and total absolute
$B$ magnitude for elliptical
galaxies in the OBEY sample (solid points) and other bright galaxies in the
same volume (open symbols). The grey band indicates our selection: a $\pm 0.15$
dex band containing galaxies with a median mass of $\log M_n = 11.45$. Most
galaxies in this mass range are in the OBEY sample. The only exceptions are
NGC\,524, NGC\,1316, and NGC\,5266 as all three are classified as S0 in
Tully (1988). However, all three galaxies have large bulges and
presumably similar surface brightness profiles as the other
14 galaxies in this mass bin.
\label{obeymos.plot}}
\end{figure*}

The relation between stellar mass and total absolute $B$ magnitude is shown in Fig.\
\ref{obeymos.plot}. Solid symbols are galaxies in the OBEY sample. Open symbols
are galaxies with $M_B^T\lesssim-21$ in the {Tully} (1988) catalog that are not
classified as elliptical galaxies, and therefore not in the OBEY sample. We note
that for one of these galaxies we adopted a different distance than is listed
in the {Tully} (1988) atlas: for NGC\,4594 (M104)
we used a distance of 9.1\,Mpc ({Jensen} {et~al.} 2003) rather than 21\,Mpc.
The grey band indicates the same selection as was used at higher redshift:
a mass bin of width $\pm 0.15$ dex and median mass determined by
Eq.\ \ref{mevo.eq} ($\log M_n = 11.45$ for $z=0$). This bin contains 14 elliptical
galaxies from the OBEY sample: NGC\,1399, NGC\,1407, NGC\,2986, IC\,3370,
NGC\,3585,  NGC\,3706, NGC\,4697, NGC\,4767, NGC\,5044, NGC\,5077,
NGC\,5813, NGC\,5846, NGC\,6861, and NGC\,6868.

\noindent
Only three galaxies have
masses near $M_n$ but are not in the OBEY sample: NGC\,524, NGC\,1316, and
NGC\,5266. All three are classified as S0 in the {Tully} (1988) atlas.
NGC\,524 is a face-on S0, but it has a velocity dispersion of 235\,\kms\ and
an effective radius of 9\,kpc ({Emsellem} {et~al.} 2007)
--- close to the average $r_e$ of the OBEY stack. NGC\,1316 is the well-known
radio galaxy Fornax A. It is a merger remnant with striking dust lanes
({Schweizer} 1980) and significant mid-IR emission ({Temi}, {Mathews}, \& {Brighenti} 2005).
Its effective radius is $\approx 8$\,kpc (e.g., {Temi} {et~al.} 2005), although 
this may be an underestimate as the galaxy has a large halo of diffuse light
(e.g., {Schweizer} 1980). NGC\,5266 has a prominent dust lane, but
can otherwise be considered an elliptical galaxy (e.g., {Varnas} {et~al.} 1987).
In summary, although the OBEY sample is not a mass-limited sample, it
misses less than 20\,\% of galaxies in the relevant mass range and there
is no indication that the galaxies that are missed have different surface
density profiles from the OBEY galaxies.

An average stack was created from the 14 OBEY galaxies.
Rather than averaging the galaxies themselves we averaged the 2D surface brightness
distributions that were measured by {Tal} {et~al.} (2009). These model images are excellent
representations of the galaxies and avoid contamination from the many neighboring stars
and galaxies.
Each galaxy was normalized to the flux inside a 24\,kpc $\times$ 24\,kpc
region centered on the galaxy (equivalent to the $10 \times 10$ pixel box
used at higher redshift). After averaging the galaxies
the flux outside $r=75$\,kpc was set to zero and
the total mass was normalized according to Eq.\,\ref{prof.eq}.

For the analysis in \S\,\ref{kin.sec} velocity dispersions of the OBEY
galaxies were obtained from the literature, using the Leda database.
They come from a variety of sources; when multiple measurements
were available we preferentially used data from
{Faber} {et~al.} (1989), {Franx}, {Illingworth}, \&  {Heckman} (1989), or {J\o{}rgensen} {et~al.} (1995). They are
indicative only as they are not necessarily measured in a
homogeneous way and do not necessarily correspond to the same
physical aperture.

\subsection{Comparison to Other Studies}

\begin{figure*}[htb]
\epsfxsize=8cm
\epsffile[-230 177 235 637]{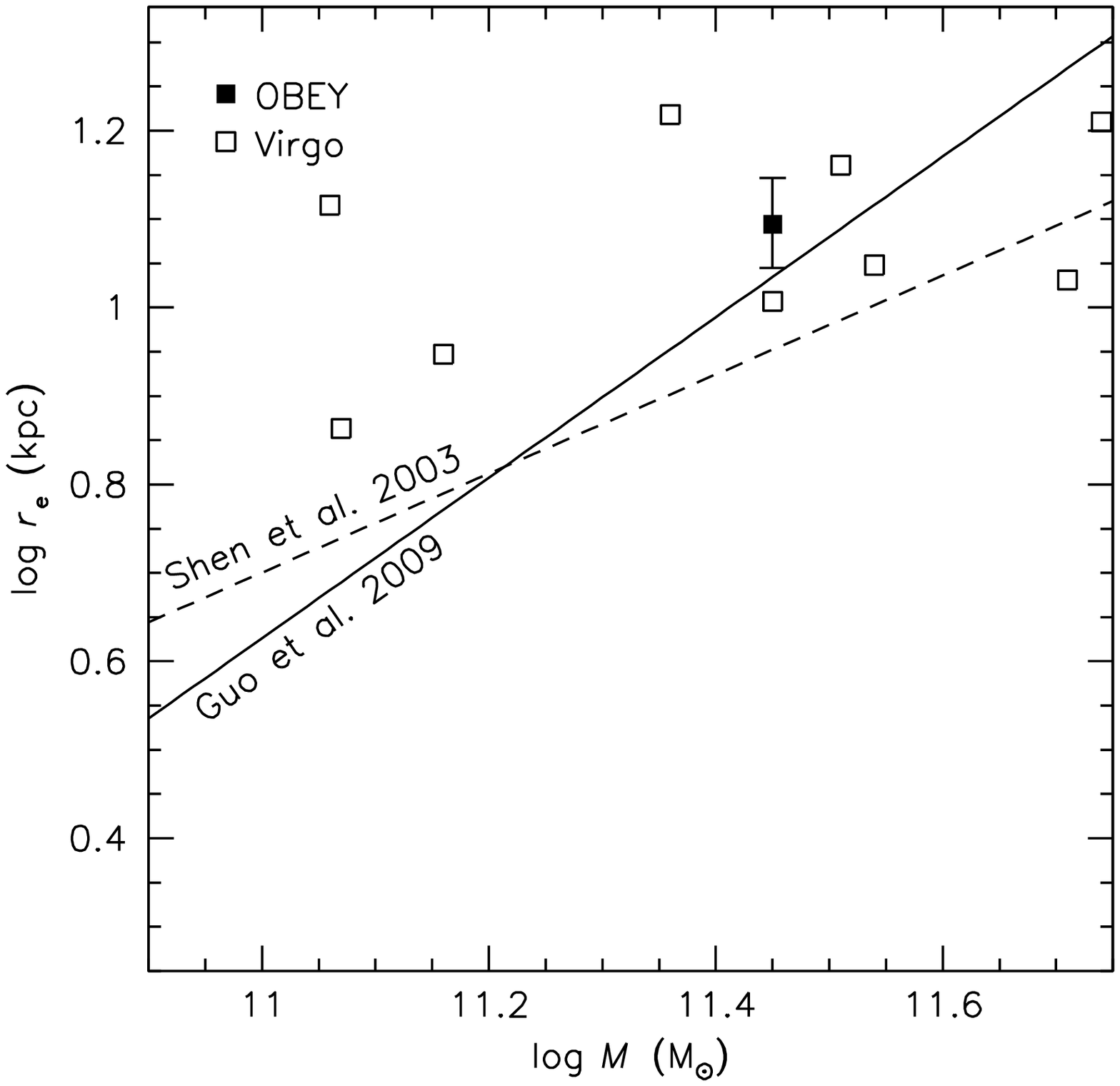}
\caption{\small Comparison of mass-size relations at $z=0$. Grey points
{\bf \em (not shown in arXiv version due to file size
restrictions)} are data
from Blanton et al.\ (2003) and Kauffmann et al.\ (2003). The dashed line is the
fit from Shen et al.\ (2003) to early-type galaxies, based on these data.
Guo et al.\ (2009) infer that Blanton et al.\ underestimated the sizes of
massive galaxies, and find a steeper relation. Open squares are elliptical
galaxies in Virgo, whose sizes were measured by Kormendy et al.\ (2009). The solid
square is our measurement from the OBEY sample.
\label{z0comp.plot}}
\end{figure*}

Here we briefly compare our datapoint at $z=0$ from the OBEY sample to results from other
recent studies. {Shen} {et~al.} (2003) determined the mass-size relation for early-type galaxies
from SDSS data, using masses from {Kauffmann} {et~al.} (2003a) and sizes from
{Blanton} {et~al.} (2003). Their relation is shown by the dashed line in Fig.\ \ref{z0comp.plot}.
Grey points are individual galaxies taken from the NYU Value Added Galaxy
Catalog (NYU-VACG) website ({Blanton} {et~al.} 2005). They are in good agreement with
the Shen et al.\ relation, as expected. The solid line shows the relation obtained
by {Guo} {et~al.} (2009) for SDSS early-type galaxies. These authors find a significantly
steeper relation then {Shen} {et~al.} (2003), possibly because {Blanton} {et~al.} (2003) underestimated
both the size and the luminosity of bright galaxies. As implied
by Eq.\ \ref{drdmfit.eq} (and as shown in Appendix A of Guo et al.\ 2009)  errors
in $r_e$ can be much larger than errors in the total luminosity, if
flux is missed at large radii.

\noindent
We also compare our datapoint to data for individual galaxies in the Virgo cluster.
{Kormendy} {et~al.} (2009) determined effective radii of elliptical galaxies in Virgo by
integrating their surface brightness profiles, using very deep and homogeneous data.
These are arguably the most accurate half-light radii for elliptical galaxies yet
measured. We determined masses for the galaxies in the {Kormendy} {et~al.} (2009) sample
in the same way as was done for the OBEY sample. Open squares in Fig.\ \ref{z0comp.plot}
indicate the  masses and sizes of the Virgo ellipticals.

The OBEY point falls very close to the relation of {Guo} {et~al.} (2009) and to the four
Virgo elliptical galaxies that have masses near $3\times 10^{11}$\,\msun. The
average values for these four galaxies are plotted in Fig.\ \ref{massradius.plot}
in the main text of the paper. The difference between the Guo et al.\ relation
and the OBEY point can easily be explained by a $0.05-0.1$ systematic difference
in $\log M$ or sample variance, as the difference
is only slightly more than $1\sigma$.

\end{appendix}

\end{document}